\documentclass[twocolumn,aps,prb,amsfonts,amsmath,amssymb,floatfix]{revtex4}

\usepackage[dvips]{graphicx}
\allowdisplaybreaks[1]

\begin{document}

\noindent
{\bf \large Quantum error correction beyond qubits}

% \author
\vskip 0.1truein
\noindent
{Takao Aoki$^{1}$, Go Takahashi$^{1,2}$, Tadashi Kajiya$^{1,2}$\\
Jun-ichi Yoshikawa$^{1,2}$, Samuel L.\ Braunstein$^{3}$,\\
Peter van Loock$^{4}$
\& Akira Furusawa$^{1,2}$}

% \affiliation
\vskip 0.05truein
\noindent
{\small
$^{1}$Department of Applied Physics, School of Engineering,
The University of Tokyo, 7-3-1 Hongo, Bunkyo-ku, Tokyo 113-8656, Japan \\
$^{2}$CREST, Japan Science and Technology (JST) Agency,
1-9-9 Yaesu, Chuo-ku, Tokyo 103-0028, Japan \\
$^{3}$Computer Science, University of York, York YO10 5DD, UK\\
$^{4}$Optical Quantum Information Theory Group,
Institute of Theoretical Physics I and Max-Planck Research Group,
Institute of Optics, Information and Photonics,
Universit\"{a}t Erlangen-N\"{u}rnberg, Staudtstr. 7/B2,
91058 Erlangen, Germany
}

% \maketitle

\vskip 0.05truein
\noindent
{\bf
Quantum computation and communication rely on the ability to manipulate
quantum states robustly and with high fidelity. Thus, some form of error
correction is needed to protect fragile quantum superposition states
from corruption by so-called decoherence noise. Indeed, the discovery of
quantum error correction (QEC) \cite{Shor95,Steane96} turned the field
of quantum information from an academic curiosity into a developing
technology. Here we present a continuous-variable experimental
implementation of a QEC code, based upon entanglement among 9
optical beams \cite{Braunstein98}. In principle, this 9-wavepacket
adaptation of Shor's original 9-qubit scheme \cite{Shor95}
allows for full quantum error correction against an arbitrary single-beam
(single-party) error.}

QEC protocols eliminate uncontrolled errors that affect fragile quantum
superposition states by encoding these quantum states into a larger,
multi-partite entangled system.  Errors occurring on a limited number of
parties will leave the entanglement intact and so the original state may
be retrieved by error syndrome recognition followed by recovery operations.
Shor proposed a concatenated quantum code to protect against arbitrary
single-qubit errors, by encoding an arbitrary single-qubit state
$|\psi\rangle =\alpha |0\rangle + \beta |1\rangle$ into nine physical
qubits
\begin{equation}
|\psi_{\rm encode}\rangle= \alpha |+,+,+\rangle + \beta |-,-,-\rangle\;,
\label{eq1}
\end{equation}
with $|\pm\rangle = (|0,0,0\rangle \pm |1,1,1\rangle)/\sqrt{2}$.
Though reminiscent of the redundant encoding in classical error correction,
the quantum code exhibits some clearly nonclassical features of which
the most significant is the presence of multi-party entanglement.
The concatenation of three-party entangled states ($|\pm\rangle$)
into nine-party states enables one to correct both bit-flip and
phase-flip errors. The latter type of error occurs only in nonclassical
states.  Remarkably, suitable error syndrome measurements would
collapse an {\it arbitrary} error (including coherent superpositions of
bit-flip and phase-flip errors) into the discrete set of only bit-flip
and/or phase-flip errors.  These discrete (Pauli) errors can be easily
reversed to recover the original state.

The continuous-variable version of Shor's 9-qubit code
\cite{Shor95,Braunstein98} is the only code to date which can be
deterministically (unconditionally) implemented
using only linear optics and sources of
entanglement.  Indeed, previous implementations of QEC were based on
qubit codes, either in liquid-state NMR or linear ion trap hardware
configurations. The liquid-state NMR experiments implemented QEC codes
with up to five physical qubits \cite{Cory98,Leung99,Knill01,Boulant05}
and in the ion trap experiment, a three-qubit code was realized
\cite{Chiaverini04}. Both configurations rely on nonlinear qubit-qubit
coupling (in the form of nearest-neighbor couplings for NMR or via the
collective vibrational mode for ion traps).  Our experiment is the first
implementation of a Shor-type code, as the preparation of nine-party
entanglement is still beyond the scope of existing non-optical approaches
and single-photon-based, optical schemes. Here, continuous-variable QEC
\cite{Braunstein98PRL,Lloyd98} is realized using squeezed states of light
and networks of beam splitters \cite{Braunstein98}. Even this optical
approach requires an optical network three times the size as that used in
earlier experiments \cite{Yonezawa04} to achieve the large-scale
multi-partite entanglement for a 9-wavepacket code.

In our scheme, as for the simplest QEC codes (whether for qubits or for
continuous variables),
a single, arbitrary error can be corrected.
Such schemes typically assume errors occur stochastically and therefore
rely on the low
frequency of multiple errors.
Stochastic error models may describe, e.g., stochastic, depolarizing
channels for qubits,
or in the continuous-variable regime \cite{PvL08,nogo}, 
free-space channels with
atmospheric fluctuations causing beam jitter,
as considered recently for various non-deterministic distillation
protocols \cite{Heersink,Dong,Schnabel,Niset}.
%Via the continuous-variable  QEC  protocols, as realized in the present
%work,
%this type of errors could be suppressed in a {\it deterministic\/} fashion
%(see supplementary material part F).
For the continuous-variable QEC protocols, as realized in the present
work, this type of error may be be suppressed in a {\it deterministic\/}
fashion (see appendix F).
The overall performance of this family of QEC codes is then only limited
by the accuracy with
which ancilla state preparation, encoding and decoding circuits, and
syndrome extraction and recovery
operations can be achieved. In the continuous-variable scheme, all these
ingredients can be
highly efficiently implemented; the finite squeezing of the auxiliary
modes being the only limitation.
This ancilla squeezing is linked with the presence of entanglement and
it also determines
whether the transfer fidelities exceed those of classical error correction
(see appendices E and F).

We begin with a description of the scheme in the limit of infinite
squeezing, where the position $x$ and momentum $p$ of a harmonic oscillator
(corresponding to a single optical mode of the light field) serve as the
conjugate pair of observables used for the encoding
\begin{equation}
|\psi_{\rm encode}\rangle= \int\,dP\, \psi(P)\,|P,P,P\rangle \;,
\end{equation}
with $|P\rangle = \frac{1}{\sqrt{\pi}}\int\,dx\,\,e^{2ix P}\,|x,x,x\rangle$,
units-free for $\hbar=\frac{1}{2}$.
Through this 9-wavepacket code an arbitrary single-mode state
$|\psi\rangle =\int\,dx\,\psi(x)|x\rangle$ is encoded into nine optical
modes. This perfectly encoded state is obtained by using eight infinitely
squeezed ancilla states. Finite squeezing of the ancillae leads to an
approximate encoding, and hence lowers the fidelity of the QEC.

% FIG 1
\begin{figure}[t]
     \begin{center}
        \includegraphics[width=0.92\linewidth]{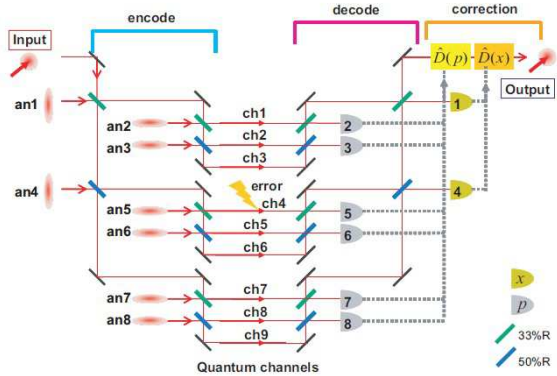}
\caption{9-wavepacket quantum error correction code\cite{Braunstein98} for
correcting an arbitrary error occurring in any one of the nine channels.
% For error syndrome recognition, the $x$
% quadrature components are measured with detectors 1 and 4, while the
% $p$ quadrature components are measured via detectors 2, 3, 5, 6, 7 and 8.
The gray dotted lines represent the classical information that is used to
compute the necessary syndrome recovery operations.}
     \end{center}
\label{fig-image-experiment}
\end{figure}

Fig.~1 shows a schematic of our realization of the 9-wavepacket code.
In the encoding stage, an input state is entangled with eight squeezed
ancillae, each corresponding to an approximate `$0$' (``blank'') state.
After an error is introduced, the states are decoded simply by inverting
the encoding. The eight ancilla modes are then measured (with
$x$-quadrature measurements performed in detectors 1 and 4 and
$p$-quadrature measurements in six other detectors), and the results
of the measurement are used for error syndrome recognition.  More precisely,
these are the results of homodyne detection applied to the ancilla modes
along their initial squeezing direction.

The encoding stage consists of two steps in order to realize the
concatenation of position and momentum codes \cite{Braunstein98}.
First, position-encoding is achieved via a {\em tritter}
${T}_{\rm in,an1,an4}$, that is two beam splitters (blue and green
in Fig.~1) acting upon the input mode and two
$x$-squeezed ancilla modes (an1 and an4 in Fig.~1).
The second step provides the momentum-encoding
via three more tritters, with six additional $p$-squeezed ancilla modes
(an2, an3, an5, an6, an7 and an8 in Fig.~1).
The overall encoding circuit becomes \cite{footnote}
\begin{equation}
{T}_{\rm an4,an7,an8}{T}_{\rm an1,an5,an6}{T}_{\rm in,an2,an3}
{T}_{\rm in,an1,an4}\;.
\end{equation}

As the decoding stage merely inverts the encoding, the eight ancilla modes
will remain all `$0$' in the absence of errors.  In the presence of an
error in any one of the nine channels, the measurement results of the
decoded ancillae will lead to non-zero components, containing sufficient
information for identifying and hence correcting the error (see appendices
for derivations and Table~\ref{tab:syndrome} for
an error-syndrome map). Similar to the qubit QEC scheme, where the
conditional state after the syndrome measurements becomes the original
input state up to some discrete Pauli errors, our conditional state
coincides with the input state up to some simple phase-space displacements.
Thus, it remains only to apply the appropriate (inverse) displacement
operations in order to correct the errors.

% TABLE 1
\begin{table}[b]
 \caption{Error syndrome measurements. LO phase: quadrature at
which the local oscillator phase of the homodyne detector is locked,
ES: equal signs, DS: different signs.}
 \begin{center}
  \begin{tabular}{|c|c|c|c|c|}
    \hline
     channel with & detectors with & LO \\
     an error & non-zero outputs & phase \\
    \hline
     1 & 1 &$x$  \\
     &  2 & $p$  \\
     \hline
      2 & 1 &$x$ \\
     &  2,3 (DS) & $p$  \\
     \hline
      3 & 1 &$x$ \\
     &  2,3 (ES) & $p$  \\
     \hline
      4 & 1,4 (DS) & $x$ \\
     &  5 & $p$  \\
     \hline
      5 & 1,4 (DS) & $x$ \\
     &  5,6 (DS) & $p$  \\
     \hline
      6 & 1,4 (DS) & $x$ \\
     &  5,6 (ES) & $p$  \\
     \hline
      7 & 1,4 (ES) & $x$ \\
     &  7 & $p$  \\
     \hline
      8 & 1,4 (ES) & $x$ \\
     &  7,8 (DS) & $p$  \\
     \hline
      9 & 1,4 (ES) &$x$ \\
     &  7,8 (ES) & $p$  \\
     \hline
     \end{tabular}
 \end{center}
\label{tab:syndrome}
\end{table}

% FIG 2
\begin{figure*}[t]
     \begin{center}
        \includegraphics[width=0.78\linewidth]{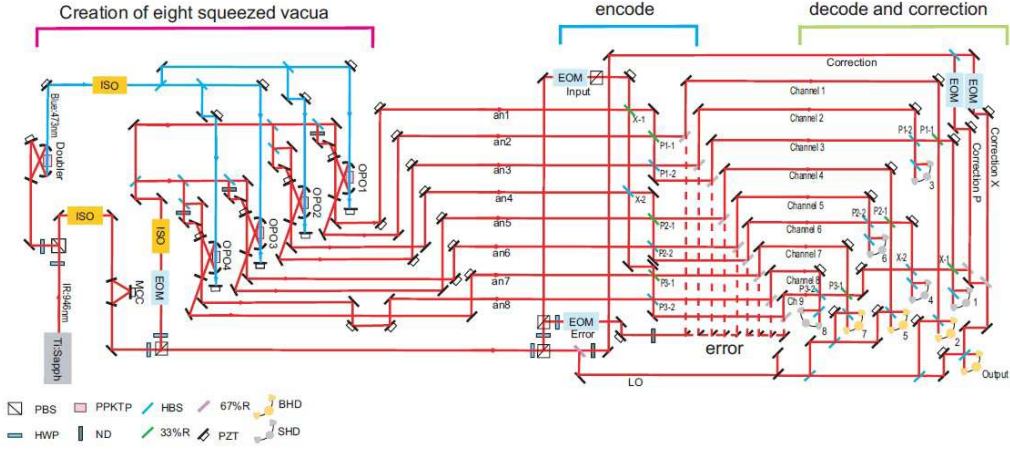}
\caption{Experimental setup of the 9-wavepacket quantum error correction;
PBS: polarization beam splitter, PPKTP: periodically poled KTiOPO$_4$,
HBS: half (symmetric) beam splitter, HWP: half wave plate,
ND: neutral density filter, PZT: piezoelectric transducer,
BHD: balanced homodyning, SHD: self-homodyning,
OPO: optical parametric oscillator, MCC: mode-cleaning cavity,
LO: local oscillator, ISO: optical isolator,
EOM: electro-optic modulator.}
     \end{center}
\label{fig-setup}
\end{figure*}

The detailed experimental setup for our 9-wavepacket QEC scheme is shown
in Fig.~2. Eight squeezed vacua are created by four optical parametric
oscillators (OPOs), which have two counter-propagating modes;
thus, every OPO creates two individual squeezed vacua.
The squeezing level of each single-mode squeezed vacuum state
corresponds to roughly 1 dB below shot noise.
For pumping the OPOs, the 2nd harmonic of a cw Ti:Sapphire laser
output is used. The syndrome measurements are performed via homodyne
detection with near-unit efficiency.

To apply a single error, a
coherent modulation is first generated in a so-called error beam using an
electro-optic modulator (EOM) (``modulated mode'').
This beam is then superimposed onto the
selected mode or channel (``target mode'')
through a high-reflectivity beam splitter
\cite{Furusawa98} with independently swept phase, resulting in a
quasi-random displacement error. The error-correcting displacement
operations (as determined by decoding and measurement) are then performed
similarly, via an EOM and a high-reflectivity beam splitter, but now with
phase locking between the modulated and target modes along either the
$x$ or $p$ axis, as appropriate.

% Table~\ref{tab:syndrome} shows error syndrome measurement outputs.
% For example, in the case of an error in channel 1,
% only detectors 1 and 2 have non-zero outputs, corresponding to $x$
% and $p$ quadrature errors, respectively. Similarly, in the case of an
% error in channel 9, only detectors 1, 4, 7, and 8 have non-zero
% outputs, where detector outcomes 1 and 4 correspond to an $x$ quadrature
% error, while the outputs have the same polarity; detector outcomes 7 and 8
% correspond to $p$ quadrature errors and the outputs have again the same
% polarity.

% FIG 3
\begin{figure}[b]
\centering
\begin{tabular}{cc}
\includegraphics[clip,scale=0.53]{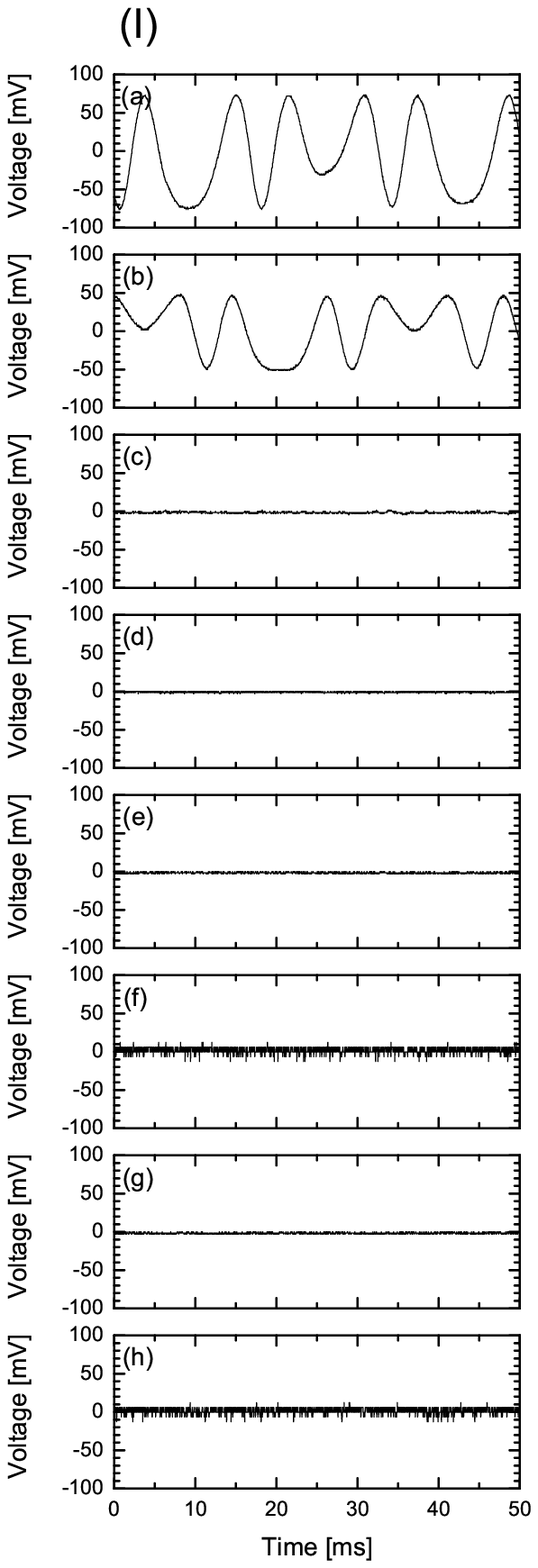} &
\includegraphics[clip,scale=0.53]{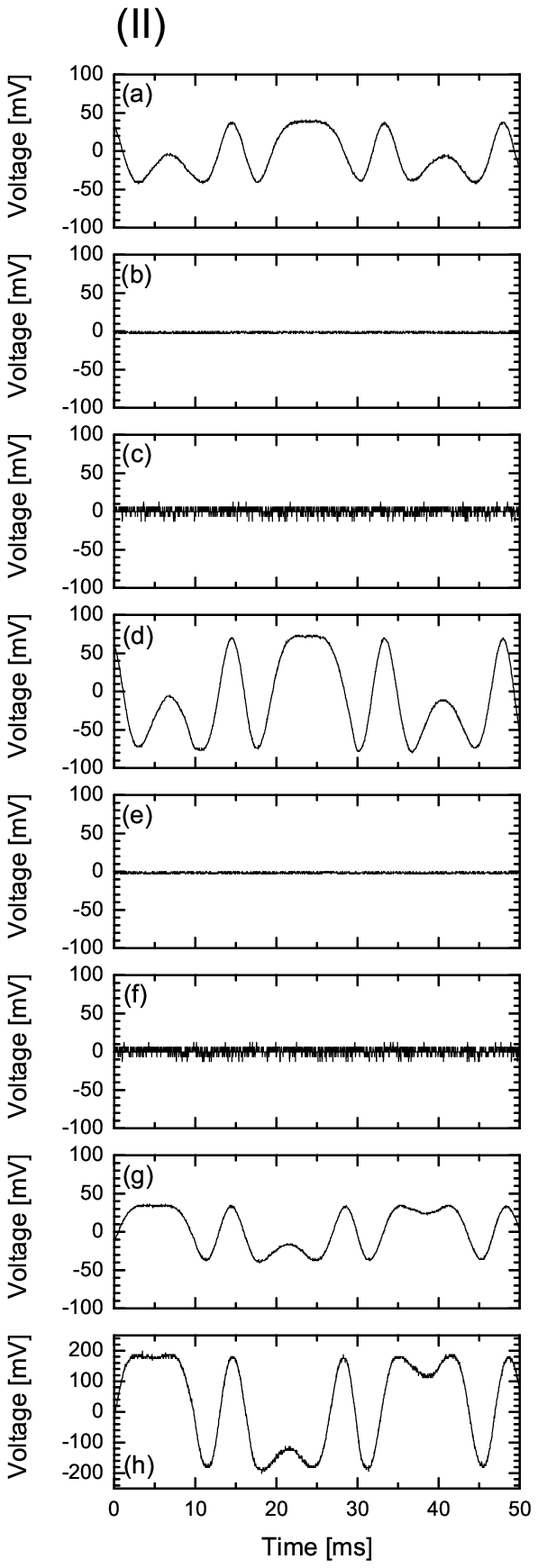}
\end{tabular}
\caption{Error syndrome measurement results. (I) A random displacement
error is imposed on channel 1. (II) A random displacement error is
imposed on channel 9. A two-channel oscilloscope is used measuring the
outputs of detectors 1 and 4, 2 and 3, 5 and 6 and 7 and 8. (a) output
signal of detector 1, (b) detector 2, (c) detector 3, (d) detector 4,
(e) detector 5, (f) detector 6, (g) detector 7, (h) detector 8.}
\end{figure}

Fig.~3 shows some examples for error syndrome measurement results. Here,
the input state is chosen to be a vacuum state. A random displacement error
in phase space is imposed on channel 1 (Fig.~3(I)) and on channel 9
(Fig.~3(II)). A two-channel oscilloscope is used to measure the outputs of
pairs of detectors (1, 4), (2, 3), (5, 6), and (7, 8). Comparing the
results of Fig.~3(I) with Table 1, one can identify an error occurring in
channel 1, since only detectors 1 and 2 have non-zero outputs. The outputs
from detectors 1 and 2 correspond to the desired $x$ and $p$ displacements,
respectively.
% Then we can feed-forward the measurement
% outcomes into $x$ and $p$ displacements via EOMs, where the outcome of
% detector 1 is used for $x$ and the one of detector 2 for $p$ displacements.
Similarly, from Fig.~3(II), we can recognize that an error has occurred
in channel 9. Here, detectors 1, 4, 7, and 8 have non-zero outputs and
the outputs of detectors 1 and 4, as well as 7 and 8 have equal signs
(distinguishing it from the case of an error in channel 8,
for which outcomes 7 and 8 have different signs).

Fig.~4 shows two examples of QEC results, comparing output states with
and without error correction, and with and without squeezing of the ancilla
modes.  In Fig.~4(I), an error was introduced in channel 1.  The local
oscillator (LO) phase of the homodyne detector was tuned to detect the
$x$ quadrature of channel 1. Similarly, in Fig.~4(II), the error was
introduced in channel 9 and the LO phase is locked to the $p$ quadrature.
For ease of experimental implementation, only the measurement outcomes of
detectors 4 and 8 were fed forward to the error correction step. In
principle, using the combined outputs of detectors 1 and 4 for $x$ and
detectors 7 and 8 for $p$ would yield even higher fidelities.

% FIG 4
\begin{figure}[t]
\centering
\begin{tabular}{cc}
\includegraphics[clip,scale=0.53]{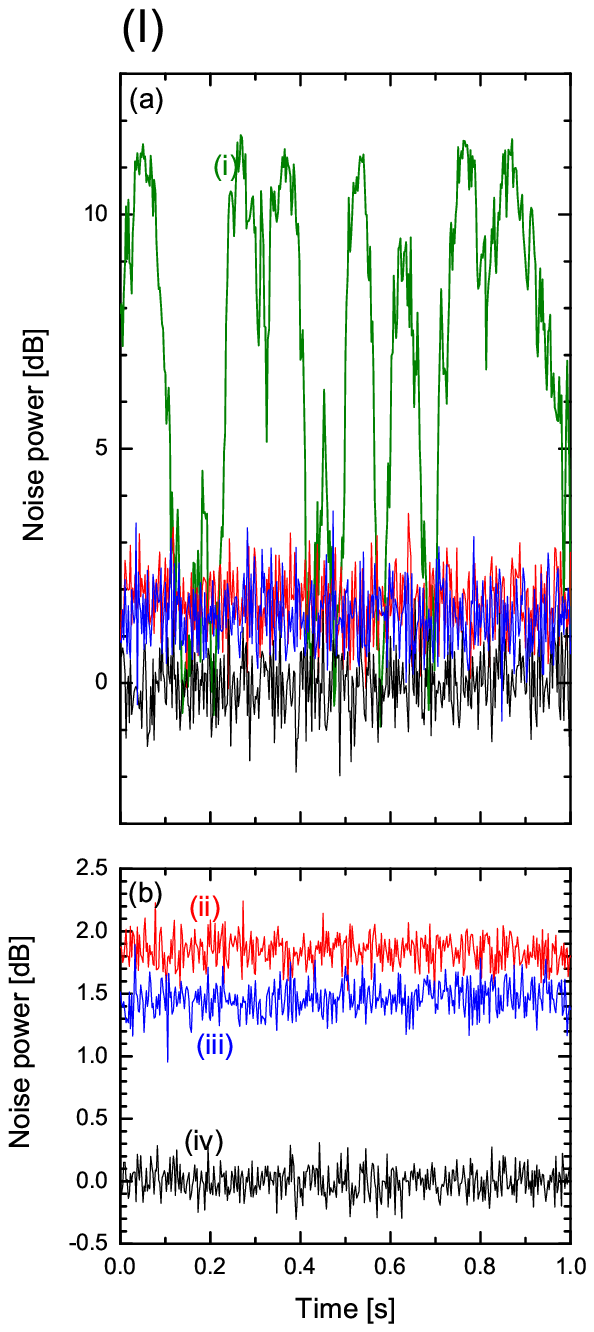} &
\includegraphics[clip,scale=0.53]{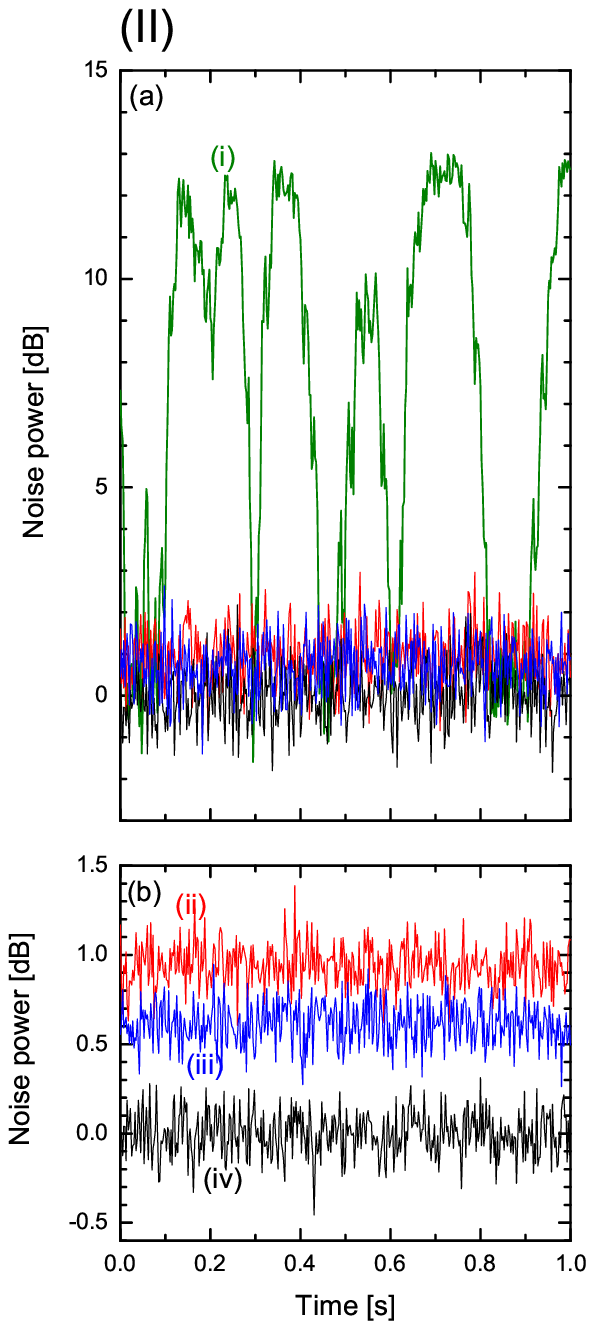}
\end{tabular}
\caption{Results of quantum error correction. (I) A random phase-space
displacement error is imposed on channel 1. The LO phase of the
homodyne detector is locked to the $x$ quadrature. (II) A random
displacement error is imposed on channel 9. The LO phase of the
homodyne detector is locked to the $p$ quadrature. In each case, four
traces are shown comparison: (i) Homodyne detector output without error
correction (no feed forward step). (ii) Error correction output without
squeezing.  (iii) Error correction output with squeezing.  (iv) shot noise
level. (a) Single scan of a spectrum analyzer with zero span mode. 2 MHz
center frequency, 30 kHz resolution band width and 300 Hz video band width.
(b) 30 times average of traces (ii-iv) above.}
\end{figure}

% Table~\ref{tab:syndrome2} summarizes the results of our QEC experiment
% with the squeezing turned on and off.
The quality of the error correction can be assessed via the fidelity
$F=\langle \psi_{\rm in} | \hat{\rho}_{\rm out} | \psi_{\rm in} \rangle$,
where $|\psi_{\rm in} \rangle$ represents the input state and
$\hat{\rho}_{\rm out}$ corresponds to the output state of the error
correction circuit \cite{Furusawa98,Braunstein01,Hammerer05}.
Here the fidelity is calculated as
\begin{equation}
F=\frac{2}{\sqrt{(1+4\langle ( \Delta \hat{x}_{\rm out})^2 \rangle)
(1+4\langle ( \Delta \hat{p}_{\rm out})^2 \rangle)}}\;,
\end{equation}
where $\hat{x}_{\rm out}$ and $\hat{p}_{\rm out}$ are quadrature
operators of the output field.
For example, in the case of an error in channel 1, the output
quadrature operators become
\begin{eqnarray}
\hat{x}_{\rm out}
&=& \hat{x}_{\rm in} - \frac{1}{\sqrt{2}} \hat{x}^{(0)}_{\rm an1}e^{-r_1}
\nonumber \\
\hat{p}_{\rm out}
&=& \hat{p}_{\rm in} - \frac{1}{\sqrt{6}} \hat{p}^{(0)}_{\rm an2}e^{-r_2}\;,
\label{eq5}
\end{eqnarray}
where $\hat{x}_{\rm in}$, $\hat{p}_{\rm in}$, $\hat{x}^{(0)}_{\rm an1}$,
and $\hat{p}^{(0)}_{\rm an2}$ are quadrature operators of the input field
and the ancilla vacuum modes, and $r_i$ are squeezing parameters for
ancilla $i$. In the ideal case of $r_i \to \infty$, unit fidelity
is obtained, with output states approaching the input states.
% $\langle ( \Delta \hat{x}_{\rm out})^2 \rangle \to
% \langle ( \hat{x}_{\rm in})^2 \rangle = \frac{1}{4}$ and
% $\langle ( \Delta \hat{p}_{\rm out})^2 \rangle \to
% \langle ( \hat{p}_{\rm in})^2 \rangle = \frac{1}{4}$.
% For finite squeezing, % the excess noise is given by
% \begin{eqnarray}
% \langle (\hat{x}_{\rm out})^2 \rangle
% &=& \langle (\hat{x}_{\rm in})^2 \rangle + \frac{1}{2} \cdot
% \frac{1}{4} e^{-2 r_1} \nonumber \\
% \langle (\hat{p}_{\rm out})^2 \rangle
% &=& \langle (\hat{p}_{\rm in})^2 \rangle + \frac{1}{6} \cdot
% \frac{1}{4} e^{-2 r_2}\;.
% \end{eqnarray}
For zero squeezing, Eq.~(\ref{eq5}) yields an excess noise of
$\frac{1}{2}$ and $\frac{1}{6}$ for the $x$ and $p$ quadratures,
corresponding to 1.76 dB and 0.67 dB of output powers, respectively
(see Table~\ref{tab:syndrome2}).
% compared to the shot noise level of $\frac{1}{4}$ (vacuum noise level)

Eq.~(4) can be used to translate the measured noise level values from
Table~\ref{tab:syndrome2} into fidelity values.  Indeed, for every
possible error introduced (in any of the channels) the fidelity after
error correction exceeds the maximum values achievable for the scheme
in the absence of ancilla squeezing. For example, for an error in
mode 1, a fidelity of 0.88$\pm$0.01 was achieved (exceeding the classical
cutoff of 0.86). Similarly, for an error in channel 9, we obtain a fidelity
of 0.86$\pm$0.01, exceeding a cutoff of 0.82. (The lower cutoff takes
into consideration that only two of the four non-zero components
are used.) The better-than-classical fidelities for errors in {\it any}
one of the nine
channels are indirect evidence of entanglement-enhanced error correction
(see appendices). By comparison,
in complete absence of any error correction, i.e.,
without reversing displacement errors
(including the zero-squeezing case; for an application of such ``classical''
error correction, see appendix F),
fidelity values under $0.007\pm0.001$ were obtained.

% TABLE 2
\begin{table}[t]
\caption{Output noise power of QEC circuit in dB, relative to the shot
noise level. Perfect error correction therefore corresponds to 0 dB.
SQV: squeezed vacua.}
 \begin{center}
  \begin{tabular}{|c|c|c|c|c|}
    \hline
     error & quadrature& output power & output power & output power \\
     on & of output & without SQV & without SQV & with SQV \\
     mode &        &      (theory) & (experiment)  & (experiment)  \\
    \hline
     1 & $x$ & 1.76 & 1.84 $\pm$ 0.12 & 1.46 $\pm$ 0.13  \\
      & $p$ & 0.67 & 0.68 $\pm$ 0.12 & 0.57 $\pm$ 0.12  \\
    \hline
     2 & $x$ & 1.76 & 1.75 $\pm$ 0.12 & 1.42 $\pm$ 0.13  \\
      & $p$ & 0.87 & 0.97 $\pm$ 0.12 & 0.72 $\pm$ 0.12  \\
    \hline
     3 & $x$ & 1.76 & 1.83 $\pm$ 0.12 & 1.41 $\pm$ 0.12  \\
      & $p$ & 0.87 & 0.92 $\pm$ 0.12 & 0.70 $\pm$ 0.12  \\
\hline
     4 & $x$ & 2.22 & 2.26 $\pm$ 0.12 & 1.67 $\pm$ 0.12  \\
      & $p$ & 0.67 & 0.73 $\pm$ 0.12 & 0.50 $\pm$ 0.12  \\
\hline
     5 & $x$ & 2.22 & 2.33 $\pm$ 0.12 & 1.79 $\pm$ 0.12  \\
      & $p$ & 0.87 & 0.88 $\pm$ 0.12 & 0.73 $\pm$ 0.13  \\
\hline
     6 & $x$ & 2.22 & 2.34 $\pm$ 0.12 & 1.77 $\pm$ 0.12  \\
      & $p$ & 0.87 & 0.87 $\pm$ 0.13 & 0.73 $\pm$ 0.13  \\
\hline
     7 & $x$ & 2.22 & 2.30 $\pm$ 0.13 & 1.72 $\pm$ 0.12  \\
      & $p$ & 0.67 & 0.69 $\pm$ 0.12 & 0.57 $\pm$ 0.12  \\
\hline
     8 & $x$ & 2.22 & 2.18 $\pm$ 0.13 & 1.79 $\pm$ 0.13  \\
      & $p$ & 0.87 & 0.84 $\pm$ 0.12 & 0.65 $\pm$ 0.12  \\
\hline
     9 & $x$ & 2.22 & 2.18 $\pm$ 0.14 & 1.82 $\pm$ 0.13  \\
      & $p$ & 0.87 & 0.94 $\pm$ 0.12 & 0.61 $\pm$ 0.12  \\
\hline
     \end{tabular}
 \end{center}
\label{tab:syndrome2}
\end{table}

In conclusion, we experimentally demonstrated a Shor-type quantum error
correction scheme based upon entanglement among nine optical beams.
The entanglement is used for deterministically generating a concatenated code,
allowing for the correction of arbitrary errors in any one of nine
communication channels. In the experiment, evidence is obtained
for an entanglement-enhanced correction of displacement errors;
a further increase of the small enhancement of the current implementation
would only require higher squeezing levels of the resource states.
Our experiment represents the first demonstration of quantum
error correction beyond qubits (and specifically for continuous variables).
The scheme may be useful for any application in which stochastic errors
occur such as free-space communication with fluctuating losses
and beam pointing errors \cite{Heersink,Dong,Schnabel,Niset}.
The ability to implement QEC in an optical network of this size represents
a significant step towards the manipulation and application of large-scale
multi-partite entanglement for quantum information processing.

This work was partly supported by SCF and GIA commissioned by the MEXT of
Japan.  PvL acknowledges the DFG for funding under the Emmy Noether
programme. AF acknowledges Y. Takeno for preparing the figures.
TA's current address is PRESTO, Japan Science and Technology Agency (JST),
Saitama, Japan.

\appendix

\section{Encoding}

Equation~(3) describes a nine-port device acting upon
the signal input mode, two
$x$-squeezed ancilla modes (``an1'' and ``an4'' in Fig.~1 of main body),
and six $p$-squeezed ancilla modes (``an2'', ``an3'', ``an5'', ``an6'',
``an7'', and ``an8'' in Fig.~1). Labeling the nine input modes
by subscripts one through nine, we obtain the output quadrature
operators of the encoded state,
\begin{eqnarray}
\hat{x}_{1} &=& \frac{1}{3}\hat{x}_{\rm in}
+ \frac{\sqrt{2}}{3}\hat{x}_{\rm an1}^{(0)}e^{-r_1}
+ \sqrt{\frac{2}{3}}\hat{x}_{\rm an2}^{(0)}e^{r_2}, \nonumber \\
\hat{p}_{1} &=& \frac{1}{3}\hat{p}_{\rm in}
+ \frac{\sqrt{2}}{3}\hat{p}_{\rm an1}^{(0)}e^{r_1}
+ \sqrt{\frac{2}{3}}\hat{p}_{\rm an2}^{(0)}e^{-r_2}, \nonumber \\
\hat{x}_{2} &=& \frac{1}{3}\hat{x}_{\rm in}
+ \frac{\sqrt{2}}{3}\hat{x}_{\rm an1}^{(0)}e^{-r_1}
- \sqrt{\frac{1}{6}}\hat{x}_{\rm an2}^{(0)}e^{r_2}
\nonumber \\
& &
+ \sqrt{\frac{1}{2}} \hat{x}_{\rm an3}^{(0)}e^{r_3}, \nonumber \\
\hat{p}_{2} &=& \frac{1}{3}\hat{p}_{\rm in}
+ \frac{\sqrt{2}}{3}\hat{p}_{\rm an1}^{(0)}e^{r_1}
- \sqrt{\frac{1}{6}}\hat{p}_{\rm an2}^{(0)}e^{-r_2}
\nonumber \\
& &
+ \sqrt{\frac{1}{2}} \hat{p}_{an3}^{(0)}e^{-r_3}, \nonumber \\
\hat{x}_{3} &=& \frac{1}{3}\hat{x}_{\rm in}
+ \frac{\sqrt{2}}{3}\hat{x}_{\rm an1}^{(0)}e^{-r_1}
- \sqrt{\frac{1}{6}}\hat{x}_{\rm an2}^{(0)}e^{r_2}
\nonumber \\
& &
- \sqrt{\frac{1}{2}} \hat{x}_{\rm an3}^{(0)}e^{r_3}, \nonumber \\
\hat{p}_{3} &=& \frac{1}{3}\hat{p}_{\rm in}
+ \frac{\sqrt{2}}{3}\hat{p}_{\rm an1}^{(0)}e^{r_1}
- \sqrt{\frac{1}{6}}\hat{p}_{\rm an2}^{(0)}e^{-r_2}
\nonumber \\
& &
- \sqrt{\frac{1}{2}} \hat{p}_{\rm an3}^{(0)}e^{-r_3}, \nonumber \\
\hat{x}_{4} &=&  \frac{1}{3}\hat{x}_{\rm in}
- \frac{1}{3\sqrt{2}}\hat{x}_{\rm an1}^{(0)}e^{-r_1}
+ \sqrt{\frac{1}{6}}\hat{x}_{\rm an4}^{(0)}e^{-r_4}
\nonumber \\
& &
+ \sqrt{\frac{2}{3}} \hat{x}_{\rm an5}^{(0)}e^{r_5}, \nonumber \\
\hat{p}_{4} &=&  \frac{1}{3}\hat{p}_{\rm in}
- \frac{1}{3\sqrt{2}}\hat{p}_{\rm an1}^{(0)}e^{r_1}
+ \sqrt{\frac{1}{6}}\hat{p}_{\rm an4}^{(0)}e^{r_4}
\nonumber \\
& &
+ \sqrt{\frac{2}{3}} \hat{p}_{\rm an5}^{(0)}e^{-r_5}, \nonumber \\
\hat{x}_{5} &=&  \frac{1}{3}\hat{x}_{\rm in}
- \frac{1}{3\sqrt{2}}\hat{x}_{\rm an1}^{(0)}e^{-r_1}
+ \sqrt{\frac{1}{6}}\hat{x}_{an4}^{(0)}e^{-r_4} \nonumber \\
& & - \sqrt{\frac{1}{6}} \hat{x}_{\rm an5}^{(0)}e^{r_5}
 + \sqrt{\frac{1}{2}}\hat{x}_{\rm an6}^{(0)}e^{r_6}, \nonumber \\
\hat{p}_{5} &=&  \frac{1}{3}\hat{p}_{\rm in}
- \frac{1}{3\sqrt{2}}\hat{p}_{\rm an1}^{(0)}e^{r_1}
+ \sqrt{\frac{1}{6}}\hat{p}_{\rm an4}^{(0)}e^{r_4}
\nonumber \\
& &
- \sqrt{\frac{1}{6}} \hat{p}_{\rm an5}^{(0)}e^{-r_5}
+ \sqrt{\frac{1}{2}}\hat{p}_{\rm an6}^{(0)}e^{-r_6}, \nonumber \\
\hat{x}_{6} &=&  \frac{1}{3}\hat{x}_{\rm in}
- \frac{1}{3\sqrt{2}}\hat{x}_{\rm an1}^{(0)}e^{-r_1}
+ \sqrt{\frac{1}{6}}\hat{x}_{\rm an4}^{(0)}e^{-r_4}
\nonumber \\
& &
- \sqrt{\frac{1}{6}} \hat{x}_{\rm an5}^{(0)}e^{r_5}
- \sqrt{\frac{1}{2}}\hat{x}_{\rm an6}^{(0)}e^{r_6}, \nonumber \\
\hat{p}_{6} &=&  \frac{1}{3}\hat{p}_{\rm in}
- \frac{1}{3\sqrt{2}}\hat{p}_{\rm an1}^{(0)}e^{r_1}
+ \sqrt{\frac{1}{6}}\hat{p}_{\rm an4}^{(0)}e^{r_4}
\nonumber \\
& &
- \sqrt{\frac{1}{6}} \hat{p}_{\rm an5}^{(0)}e^{-r_5}
- \sqrt{\frac{1}{2}}\hat{p}_{\rm an6}^{(0)}e^{-r_6}, \nonumber \\
 \hat{x}_{7} &=&  \frac{1}{3}\hat{x}_{\rm in}
- \frac{1}{3\sqrt{2}}\hat{x}_{\rm an1}^{(0)}e^{-r_1}
\nonumber \\
& &
- \sqrt{\frac{1}{6}}\hat{x}_{\rm an4}^{(0)}e^{-r_4}
+ \sqrt{\frac{2}{3}} \hat{x}_{\rm an7}^{(0)}e^{r_7}, \nonumber \\
\hat{p}_{7} &=&  \frac{1}{3}\hat{p}_{\rm in}
- \frac{1}{3\sqrt{2}}\hat{p}_{\rm an1}^{(0)}e^{r_1}
- \sqrt{\frac{1}{6}}\hat{p}_{\rm an4}^{(0)}e^{r_4}
\nonumber \\
& &
+ \sqrt{\frac{2}{3}} \hat{p}_{\rm an7}^{(0)}e^{-r_7}, \nonumber \\
\hat{x}_{8} &=&  \frac{1}{3}\hat{x}_{\rm in}
- \frac{1}{3\sqrt{2}}\hat{x}_{\rm an1}^{(0)}e^{-r_1}
- \sqrt{\frac{1}{6}}\hat{x}_{\rm an4}^{(0)}e^{-r_4}
\nonumber \\
& &
- \sqrt{\frac{1}{6}} \hat{x}_{\rm an7}^{(0)}e^{r_7}
+ \sqrt{\frac{1}{2}}\hat{x}_{\rm an8}^{(0)}e^{r_8}, \nonumber \\
\hat{p}_{8} &=&  \frac{1}{3}\hat{p}_{\rm in}
- \frac{1}{3\sqrt{2}}\hat{p}_{\rm an1}^{(0)}e^{r_1}
- \sqrt{\frac{1}{6}}\hat{p}_{\rm an4}^{(0)}e^{r_4}
\nonumber \\
& &
- \sqrt{\frac{1}{6}} \hat{p}_{\rm an7}^{(0)}e^{-r_7}
+ \sqrt{\frac{1}{2}}\hat{p}_{\rm an8}^{(0)}e^{-r_8}, \nonumber \\
\hat{x}_{9} &=&  \frac{1}{3}\hat{x}_{\rm in}
- \frac{1}{3\sqrt{2}}\hat{x}_{\rm an1}^{(0)}e^{-r_1}
- \sqrt{\frac{1}{6}}\hat{x}_{\rm an4}^{(0)}e^{-r_4}
\nonumber \\
& &
- \sqrt{\frac{1}{6}} \hat{x}_{\rm an7}^{(0)}e^{r_7}
- \sqrt{\frac{1}{2}}\hat{x}_{\rm an8}^{(0)}e^{r_8}, \nonumber \\
\hat{p}_{9} &=&  \frac{1}{3}\hat{p}_{\rm in}
- \frac{1}{3\sqrt{2}}\hat{p}_{\rm an1}^{(0)}e^{r_1}
- \sqrt{\frac{1}{6}}\hat{p}_{\rm an4}^{(0)}e^{r_4}
\nonumber \\
& &
- \sqrt{\frac{1}{6}} \hat{p}_{\rm an7}^{(0)}e^{-r_7}
- \sqrt{\frac{1}{2}}\hat{p}_{\rm an8}^{(0)}e^{-r_8}.
\label{eq-encode-all}
\end{eqnarray}
Note that with respect to these subscripts,
eq.~(3) can be expressed by
${T}_{789}{T}_{456}{T}_{123}
{T}_{147}$ for modes 1 (signal input), 2 (``an2''),
3 (``an3''), 4  (``an1''), 5  (``an5''), 6  (``an6''),
7  (``an4''), 8  (``an7''), and 9  (``an8'').

The encoded state exhibits the following quadrature quantum correlations
in the case of nonzero squeezing,
\begin{eqnarray}\label{eq-th2-4}
\hat{x}_1 + \hat{x}_2 + \hat{x}_3 -&&
\!\!\!\!\!\!\!\!\!\!\!( \hat{x}_4 + \hat{x}_5 + \hat{x}_6 )\nonumber \\
&=& \frac{3}{\sqrt{2}}{\hat{x}_{\rm an1}^{(0)}}e^{-r_1}
-\sqrt{\frac{3}{2}}{\hat{x}_{\rm an4}^{(0)}}e^{-r_4},\nonumber \\
\hat{x}_4 + \hat{x}_5 + \hat{x}_6 -&&
\!\!\!\!\!\!\!\!\!\!\!( \hat{x}_7 + \hat{x}_8 + \hat{x}_9 )\nonumber \\
&=& \sqrt{6}{\hat{x}_{\rm an4}^{(0)}}e^{-r_4},\nonumber \\
\hat{p}_1-\hat{p}_2
&=& \sqrt{\frac{3}{2}}{\hat{p}_{\rm an2}^{(0)}}e^{-r_2}
-\frac{1}{\sqrt{2}}{\hat{p}_{\rm an3}^{(0)}}e^{-r_3},\nonumber \\
\hat{p}_2-\hat{p}_3
&=& \sqrt{2}{\hat{p}_{\rm an3}^{(0)}}e^{-r_3},\nonumber \\
\hat{p}_4-\hat{p}_5
&=& \sqrt{\frac{3}{2}}{\hat{p}_{\rm an5}^{(0)}}e^{-r_5}
-\frac{1}{\sqrt{2}}{\hat{p}_{\rm an6}^{(0)}}e^{-r_6},\nonumber \\
\hat{p}_5-\hat{p}_6
&=& \sqrt{2}{\hat{p}_{\rm an6}^{(0)}}e^{-r_6},\nonumber \\
\hat{p}_7-\hat{p}_8
&=& \sqrt{\frac{3}{2}}{\hat{p}_{\rm an7}^{(0)}}e^{-r_7}
-\frac{1}{\sqrt{2}}{\hat{p}_{\rm an8}^{(0)}}e^{-r_8},\nonumber \\
\hat{p}_8-\hat{p}_9
&=& \sqrt{2}{\hat{p}_{\rm an8}^{(0)}}e^{-r_8}.\nonumber \\
\end{eqnarray}
In the limit $r_{1-8} \to \infty$,
the quadrature operators become perfectly correlated,
\begin{eqnarray}
\hat{x}_1 + \hat{x}_2 + \hat{x}_3
&=&
\hat{x}_4 + \hat{x}_5 + \hat{x}_6
=
\hat{x}_7 + \hat{x}_8 + \hat{x}_9,
\nonumber \\
\hat{p}_1 &=& \hat{p}_2 = \hat{p}_3,
\nonumber \\
\hat{p}_4 &=& \hat{p}_5 = \hat{p}_6,
\nonumber \\
\hat{p}_7 &=& \hat{p}_8 = \hat{p}_9.
\label{stabilizer}
\end{eqnarray}
These correlations are analogous expressions
to the eight stabilizer conditions of the Shor qubit code
(where for continuous variables, Pauli operators are
replaced by Weyl-Heisenberg phase-space operators).
%In this case, the state is fully entangled.
%Even with non-zero squeezing, there exists multipartite
% entanglement \cite{vanLoock03}.
%Moreover, the combination of quadrature operators shown
%in the left-hand side of eq.~(\ref{eq-th2-4}) are to be
%  measured at detectors 1-8,
%respectively, and in the case of no error, they are squeezed vacua.
Note that these correlations hold for {\it any} signal input state, i.e.,
for any resulting ``code words'', again similar to the stabilizer conditions
for qubits. In order to obtain a sufficient set of entanglement witnesses
for verifying a fully inseparable nine-party state, additional
quadrature correlations must be considered; these extra correlations
are expressed in terms of the ``logical'' quadratures in the code space
which depend also on the signal state (see appendix E).

\section{Decoding and Correction}

Random phase fluctuations are transferred onto one selected beam
of the encoded state, leading to random phase-space displacements of one of the
nine optical modes. This effect can be described by adding error quadrature operators
to the corresponding mode $k$, $\lambda_k \hat{x}_k^{e}$ and $\lambda_k \hat{p}_k^{e}$,
where the parameter $\lambda_k$ will be set to one for the single mode
of the noisy quantum channel and otherwise chosen to be zero.
After the decoding step, ${T}_{147}^{-1}{T}_{123}^{-1}{T}_{456}^{-1}
{T}_{789}^{-1}$, the outgoing quadrature operators become
\begin{eqnarray}
\hat{x}_1'
&=&
\hat{x}_{\rm in} + \frac{1}{3}\sum_{k=1}^{9}\lambda_k \hat{x}_k^{e},\nonumber \\
\hat{p}_1'
&=&
\hat{p}_{\rm in} + \frac{1}{3}\sum_{k=1}^{9}\lambda_k \hat{p}_k^{e},\nonumber \\
\hat{x}_2'
&=&
\hat{x}_{\rm an2}^{(0)}e^{r_2} + \sqrt{\frac{2}{3}}
\lambda_1 \hat{x}_1^{e} - \frac{1}{\sqrt{6}}\left(\lambda_2 \hat{x}_2^{e}
+ \lambda_3 \hat{x}_3^{e}\right),\nonumber \\
\hat{p}_2'
&=&
\hat{p}_{\rm an2}^{(0)}e^{-r_2} + \sqrt{\frac{2}{3}}
\lambda_1 \hat{p}_1^{e} - \frac{1}{\sqrt{6}}\left(\lambda_2 \hat{p}_2^{e}
+ \lambda_3 \hat{p}_3^{e}\right),\nonumber \\
\hat{x}_3'
&=&
\hat{x}_{\rm an3}^{(0)}e^{r_3} +
\frac{1}{\sqrt{2}}\left(\lambda_2 \hat{x}_2^{e}
- \lambda_3 \hat{x}_3^{e}\right),\nonumber \\
\hat{p}_3'
&=&
\hat{p}_{\rm an3}^{(0)}e^{-r_3} +
\frac{1}{\sqrt{2}}\left(\lambda_2 \hat{p}_2^{e}
- \lambda_3 \hat{p}_3^{e}\right),\nonumber \\
\hat{x}_4'
&=&
\hat{x}_{\rm an1}^{(0)}e^{-r_1} +
\frac{\sqrt{2}}{3}\sum_{k=1}^{3}\lambda_k \hat{x}_k^{e}
-\frac{1}{\sqrt{18}}\sum_{k=4}^{9}\lambda_k \hat{x}_k^{e},\nonumber \\
\hat{p}_4'
&=&
\hat{p}_{\rm an1}^{(0)}e^{r_1} +
\frac{\sqrt{2}}{3}\sum_{k=1}^{3}\lambda_k \hat{p}_k^{e}
-\frac{1}{\sqrt{18}}\sum_{k=4}^{9}\lambda_k \hat{p}_k^{e},\nonumber \\
\hat{x}_5'
&=&
\hat{x}_{\rm an5}^{(0)}e^{r_5} + \sqrt{\frac{2}{3}}
\lambda_4 \hat{x}_4^{e} - \frac{1}{\sqrt{6}}\left(\lambda_5 \hat{x}_5^{e}
+ \lambda_6 \hat{x}_6^{e}\right),\nonumber \\
\hat{p}_5'
&=&
\hat{p}_{\rm an5}^{(0)}e^{-r_5} + \sqrt{\frac{2}{3}}
\lambda_4 \hat{p}_4^{e} - \frac{1}{\sqrt{6}}\left(\lambda_5 \hat{p}_5^{e}
+ \lambda_6 \hat{p}_6^{e}\right),\nonumber \\
\hat{x}_6'
&=&
\hat{x}_{\rm an6}^{(0)}e^{r_6} +
\frac{1}{\sqrt{2}}\left(\lambda_5 \hat{x}_5^{e}
- \lambda_6 \hat{x}_6^{e}\right),\nonumber \\
\hat{p}_6'
&=&
\hat{p}_{\rm an6}^{(0)}e^{-r_6} +
\frac{1}{\sqrt{2}}\left(\lambda_5 \hat{p}_5^{e}
- \lambda_6 \hat{p}_6^{e}\right),\nonumber \\
\hat{x}_7'
&=&
\hat{x}_{\rm an4}^{(0)}e^{-r_4} +
\frac{1}{\sqrt{6}}\left(
\sum_{k=4}^{6}\lambda_k \hat{x}_k^{e}
- \sum_{k=7}^{9}\lambda_k \hat{x}_k^{e}\right),\nonumber \\
\hat{p}_7'
&=&
\hat{p}_{\rm an4}^{(0)}e^{r_4} +
\frac{1}{\sqrt{6}}\left(
\sum_{k=4}^{6}\lambda_k \hat{p}_k^{e}
- \sum_{k=7}^{9}\lambda_k \hat{p}_k^{e}\right),\nonumber \\
\hat{x}_8'
&=&
\hat{x}_{\rm an7}^{(0)}e^{r_7} + \sqrt{\frac{2}{3}}
\lambda_7 \hat{x}_7^{e} - \frac{1}{\sqrt{6}}\left(\lambda_8 \hat{x}_8^{e}
+ \lambda_9 \hat{x}_9^{e}\right),\nonumber \\
\hat{p}_8'
&=&
\hat{p}_{\rm an7}^{(0)}e^{-r_7} + \sqrt{\frac{2}{3}}
\lambda_7 \hat{p}_7^{e} - \frac{1}{\sqrt{6}}\left(\lambda_8 \hat{p}_8^{e}
+ \lambda_9 \hat{p}_9^{e}\right),\nonumber \\
\hat{x}_9'
&=&
\hat{x}_{\rm an8}^{(0)}e^{r_8} +
\frac{1}{\sqrt{2}}\left(\lambda_8 \hat{x}_8^{e}
- \lambda_9 \hat{x}_9^{e}\right),\nonumber \\
\hat{p}_9'
&=&
\hat{p}_{\rm an8}^{(0)}e^{-r_8} +
\frac{1}{\sqrt{2}}\left(\lambda_8 \hat{p}_8^{e}
- \lambda_9 \hat{p}_9^{e}\right).\nonumber \\
\label{eq-out-er1}
\end{eqnarray}
Modes two through nine are measured via suitable
homodyne detectors, i.e., the local oscillator phase is
adjusted to detect those quadratures which are quiet if
there was no error.
After the corresponding feedforward operations
on the first mode, the signal input state will be recovered
in mode 1 up to the finite
squeezing from the ancilla modes.

For example, in the case of an error transferred onto mode 1,
$\lambda_k = \delta_{k1}$,
\begin{eqnarray}
\hat{x}_1'
&=&
\hat{x}_{\rm in} + \frac{1}{3}\hat{x}_1^{e},\nonumber \\
\hat{p}_1'
&=&
\hat{p}_{\rm in} + \frac{1}{3}\hat{p}_1^{e},\nonumber \\
\end{eqnarray}
only for detectors 1 and 2 (see Fig.~2 of main body), measuring $\hat{x}_4'$
(position of ``an1'') and $\hat{p}_2'$ (momentum of ``an2''), respectively,
results clearly different from
zero (coming from the error) are obtained.
All the remaining detectors show results around zero.
In order to correct the error, mode 1 is displaced according to
\begin{eqnarray}
\hat{x}_1'
&\to & \hat{x}_1' - \frac{1}{\sqrt{2}}\hat{x}_4',
\nonumber \\
\hat{p}_1'
&\to & \hat{p}_1' - \frac{1}{\sqrt{6}}\hat{p}_2',
\end{eqnarray}
leading to
\begin{eqnarray}
   \hat{x}_{\rm out} &=&
\hat{x}_{\rm in} - \frac{1}{\sqrt{2}}\hat{x}_{\rm an1}^{(0)}e^{-r_1}, \nonumber \\
   \hat{p}_{\rm out} &=&
\hat{p}_{\rm in} - \frac{1}{\sqrt{6}}\hat{p}_{\rm an2}^{(0)}e^{-r_2},
\end{eqnarray}
using eqs.~(\ref{eq-out-er1}).

Similarly, in the case of an error transferred onto mode 9,
$\lambda_k = \delta_{k9}$, we have
\begin{eqnarray}
\hat{x}_1'
&=&
\hat{x}_{\rm in} + \frac{1}{3}\hat{x}_9^{e},\nonumber \\
\hat{p}_1'
&=&
\hat{p}_{\rm in} + \frac{1}{3}\hat{p}_9^{e}.\nonumber \\
\end{eqnarray}
Now the only nonzero outputs occur at detectors 1 and 4 (Fig.~2),
measuring $\hat{x}_4'$
(position of ``an1'') and $\hat{x}_7'$ (position of ``an4''), respectively, and
at detectors 7 and 8, measuring $\hat{p}_8'$
(momentum of ``an7'') and $\hat{p}_9'$ (momentum of ``an8''), respectively.
Possible correction displacements are
\begin{eqnarray}
\hat{x}_1'
&\to & \hat{x}_1' + \sqrt{2}\hat{x}_4',
\nonumber \\
\hat{x}_1'
&\to & \hat{x}_1' + \sqrt{\frac{2}{3}}\hat{x}_7',
\end{eqnarray}
for $x$, and
\begin{eqnarray}
\hat{p}_1'
&\to & \hat{p}_1' + \sqrt{\frac{2}{3}}\hat{p}_8',
\nonumber \\
\hat{p}_1'
&\to & \hat{p}_1' + \frac{\sqrt{2}}{3}\hat{p}_9',
\end{eqnarray}
for $p$.
These corrections result in the output quadratures
\begin{eqnarray}
\hat{x}_{\rm out}
&=&
\hat{x}_{\rm in} + \sqrt{2} \hat{x}^{(0)}_{\rm an1}e^{-r_1},
\nonumber \\
\hat{x}_{\rm out}
&=&
\hat{x}_{\rm in} + \sqrt{\frac{2}{3}} \hat{x}^{(0)}_{\rm an4}e^{-r_4},
\nonumber \\
\hat{p}_{\rm out}
&=&
\hat{p}_{\rm in} + \sqrt{\frac{2}{3}} \hat{p}^{(0)}_{\rm an7}e^{-r_7},
\nonumber \\
\hat{p}_{\rm out}
&=&
\hat{p}_{\rm in} + \frac{\sqrt{2}}{3} \hat{p}^{(0)}_{\rm an8}e^{-r_8},
\end{eqnarray}
always nearly recovering the signal input state.

Similar calculations yield the quadrature operators
for the output state of mode 1 after the error correction protocol
in the case of an error on modes two through eight;
for an error on mode 2,
\begin{eqnarray}
\hat{x}_{\rm out,det1}
&=&
\hat{x}_{\rm in} - \frac{1}{\sqrt{2}} \hat{x}^{(0)}_{\rm an1}e^{-r_1},
\nonumber \\
\hat{p}_{\rm out,det2}
&=&
\hat{p}_{\rm in} + \sqrt{\frac{2}{3}} \hat{p}^{(0)}_{\rm an2}e^{-r_2},
\nonumber \\
\hat{p}_{\rm out,det3}
&=&
\hat{p}_{\rm in} - \frac{\sqrt{2}}{3} \hat{p}^{(0)}_{\rm an3}e^{-r_3},
\end{eqnarray}
for an error on mode 3,
\begin{eqnarray}
\hat{x}_{\rm out,det1}
&=&
\hat{x}_{\rm in} - \frac{1}{\sqrt{2}} \hat{x}^{(0)}_{\rm an1}e^{-r_1},
\nonumber \\
\hat{p}_{\rm out,det2}
&=&
\hat{p}_{\rm in} + \sqrt{\frac{2}{3}} \hat{p}^{(0)}_{\rm an2}e^{-r_2},
\nonumber \\
\hat{p}_{\rm out,det3}
&=&
\hat{p}_{\rm in} + \frac{\sqrt{2}}{3} \hat{p}^{(0)}_{\rm an3}e^{-r_3},
\end{eqnarray}
for an error on mode 4,
\begin{eqnarray}
\hat{x}_{\rm out,det1}
&=&
\hat{x}_{\rm in} + \sqrt{2} \hat{x}^{(0)}_{\rm an1}e^{-r_1},
\nonumber \\
\hat{x}_{\rm out,det4}
&=&
\hat{x}_{\rm in} - \sqrt{\frac{2}{3}} \hat{x}^{(0)}_{\rm an4}e^{-r_4},
\nonumber \\
\hat{p}_{\rm out,det5}
&=&
\hat{p}_{\rm in} - \frac{1}{\sqrt{6}} \hat{p}^{(0)}_{\rm an5}e^{-r_5},
\end{eqnarray}
for an error on mode 5,
\begin{eqnarray}
\hat{x}_{\rm out,det1}
&=&
\hat{x}_{\rm in} + \sqrt{2} \hat{x}^{(0)}_{\rm an1}e^{-r_1},
\nonumber \\
\hat{x}_{\rm out,det4}
&=&
\hat{x}_{\rm in} - \sqrt{\frac{2}{3}} \hat{x}^{(0)}_{\rm an4}e^{-r_4},
\nonumber \\
\hat{p}_{\rm out,det5}
&=&
\hat{p}_{\rm in} - \sqrt{\frac{2}{3}} \hat{p}^{(0)}_{\rm an5}e^{-r_5},
\nonumber \\
\hat{p}_{\rm out,det6}
&=&
\hat{p}_{\rm in} - \frac{\sqrt{2}}{3} \hat{p}^{(0)}_{\rm an6}e^{-r_6},
\end{eqnarray}
for an error on mode 6,
\begin{eqnarray}
\hat{x}_{\rm out,det1}
&=&
\hat{x}_{\rm in} + \sqrt{2} \hat{x}^{(0)}_{\rm an1}e^{-r_1},
\nonumber \\
\hat{x}_{\rm out,det4}
&=&
\hat{x}_{\rm in} - \sqrt{\frac{2}{3}} \hat{x}^{(0)}_{\rm an4}e^{-r_4},
\nonumber \\
\hat{p}_{\rm out,det5}
&=&
\hat{p}_{\rm in} + \sqrt{\frac{2}{3}} \hat{p}^{(0)}_{\rm an5}e^{-r_5},
\nonumber \\
\hat{p}_{\rm out,det6}
&=&
\hat{p}_{\rm in} + \frac{\sqrt{2}}{3} \hat{p}^{(0)}_{\rm an6}e^{-r_6},
\end{eqnarray}
for an error on mode 7,
\begin{eqnarray}
\hat{x}_{\rm out,det1}
&=&
\hat{x}_{\rm in} + \sqrt{2} \hat{x}^{(0)}_{\rm an1}e^{-r_1},
\nonumber \\
\hat{x}_{\rm out,det4}
&=&
\hat{x}_{\rm in} + \sqrt{\frac{2}{3}} \hat{x}^{(0)}_{\rm an4}e^{-r_4},
\nonumber \\
\hat{p}_{\rm out,det7}
&=&
\hat{p}_{\rm in} - \frac{1}{\sqrt{6}} \hat{p}^{(0)}_{\rm an7}e^{-r_7},
\end{eqnarray}
for an error on mode 8,
\begin{eqnarray}
\hat{x}_{\rm out,det1}
&=&
\hat{x}_{\rm in} + \sqrt{2} \hat{x}^{(0)}_{\rm an1}e^{-r_1},
\nonumber \\
\hat{x}_{\rm out,det4}
&=&
\hat{x}_{\rm in} + \sqrt{\frac{2}{3}} \hat{x}^{(0)}_{\rm an4}e^{-r_4},
\nonumber \\
\hat{p}_{\rm out,det7}
&=&
\hat{p}_{\rm in} + \sqrt{\frac{2}{3}} \hat{p}^{(0)}_{\rm an7}e^{-r_7},
\nonumber \\
\hat{p}_{\rm out,det8}
&=&
\hat{p}_{\rm in} - \frac{\sqrt{2}}{3} \hat{p}^{(0)}_{\rm an8}e^{-r_8}.
\end{eqnarray}
The additional subscripts ``det1'', etc., indicate which detector outcomes
are used for the correction displacements. These detectors (see Fig.~2)
measure the quadratures $\hat{x}_4'$ (``det1''), $\hat{p}_2'$ (``det2''),
$\hat{p}_3'$ (``det3''), $\hat{x}_7'$ (``det4''), $\hat{p}_5'$ (``det5''),
$\hat{p}_6'$ (``det6''), $\hat{p}_8'$ (``det7''), and $\hat{p}_9'$ (``det8'').

Because of the freedom in choosing the correction displacements,
there is always an optimal feedforward operation. For example,
in the case of an error on mode 2,
\begin{eqnarray}
\hat{x}_{\rm out}
&=&
\hat{x}_{\rm in} - \frac{1}{\sqrt{2}} \hat{x}^{(0)}_{\rm an1}e^{-r_1},
\nonumber \\
\hat{p}_{\rm out,det2}
&=&
\hat{p}_{\rm in} + \sqrt{\frac{2}{3}} \hat{p}^{(0)}_{\rm an2}e^{-r_2},
\nonumber \\
\hat{p}_{\rm out,det3}
&=&
\hat{p}_{\rm in} - \frac{\sqrt{2}}{3} \hat{p}^{(0)}_{\rm an3}e^{-r_3},
\end{eqnarray}
we obtain the following excess noise for the output state,
\begin{eqnarray}
\langle (\hat{x}_{\rm out})^2 \rangle
&=&
\langle (\hat{x}_{\rm in})^2 \rangle + \frac{1}{2} \cdot
\frac{1}{4} e^{-2 r_1},
\nonumber \\
\langle (\hat{p}_{\rm out,det2})^2 \rangle
&=&
\langle (\hat{p}_{\rm in})^2 \rangle + \frac{2}{3} \cdot
\frac{1}{4} e^{-2 r_2},
\nonumber \\
\langle (\hat{p}_{\rm out,det3})^2 \rangle
&=&
\langle (\hat{p}_{\rm in})^2 \rangle + \frac{2}{9} \cdot
\frac{1}{4} e^{-2 r_3}.
\end{eqnarray}
However, for $r_2=r_3=r$, the optimal feedforward operation leads to
\begin{equation}
\hat{p}_{\rm out,opt}
=
\hat{p}_{\rm in} + \frac{1}{2\sqrt{6}} \hat{p}^{(0)}_{\rm an2}e^{-r}
- \frac{\sqrt{2}}{4} \hat{p}^{(0)}_{\rm an3}e^{-r},
\end{equation}
corresponding to
\begin{equation}
\langle (\hat{p}_{\rm out,opt})^2 \rangle
=
\langle (\hat{p}_{\rm in})^2 \rangle + \frac{1}{6} \cdot
\frac{1}{4} e^{-2r},
\end{equation}
which is the same as for the case of an error on mode 1.
For unequal squeezing, $r_2 \neq r_3$,
the optimal feedforward depends on the squeezing values.
Therefore, in the current experiment, we use only the output of detector 3 for the feedforward.
Table~\ref{tab-det-feedforward} shows which outputs of the homodyne detectors
are used for error correction.

\begin{table}[htbp]
 \caption{Homodyne detector outputs for feedforward in the current experiments.}
 \begin{center}
  \begin{tabular}{|c|c|c|c|c|}
    \hline
     channel with & quadrature & detectors for  \\
     an error &  & feedforward  \\
    \hline
     1  &$x$ & 1 \\
     &   $p$ & 2 \\
     \hline
      2 & $x$ & 1\\
     &  $p$ & 3    \\
     \hline
      3 &$x$&1 \\
     &  $p$&3  \\
     \hline
      4 &  $x$&4 \\
     &  $p$ &5 \\
     \hline
      5 &  $x$&4 \\
     &  $p$ &6 \\
     \hline
      6 & $x$&4 \\
     &  $p$ &6 \\
     \hline
      7 & $x$ &4\\
     &  $p$&7  \\
     \hline
      8 & $x$&4 \\
     & $p$ &8 \\
     \hline
      9 &$x$ &4\\
     & $p$ &8 \\
     \hline
     \end{tabular}
 \end{center}
\end{table}\label{tab-det-feedforward}

\section{Results of error syndrome measurements}

\begin{figure*}[t]
\centering
\includegraphics[clip,scale=0.5]{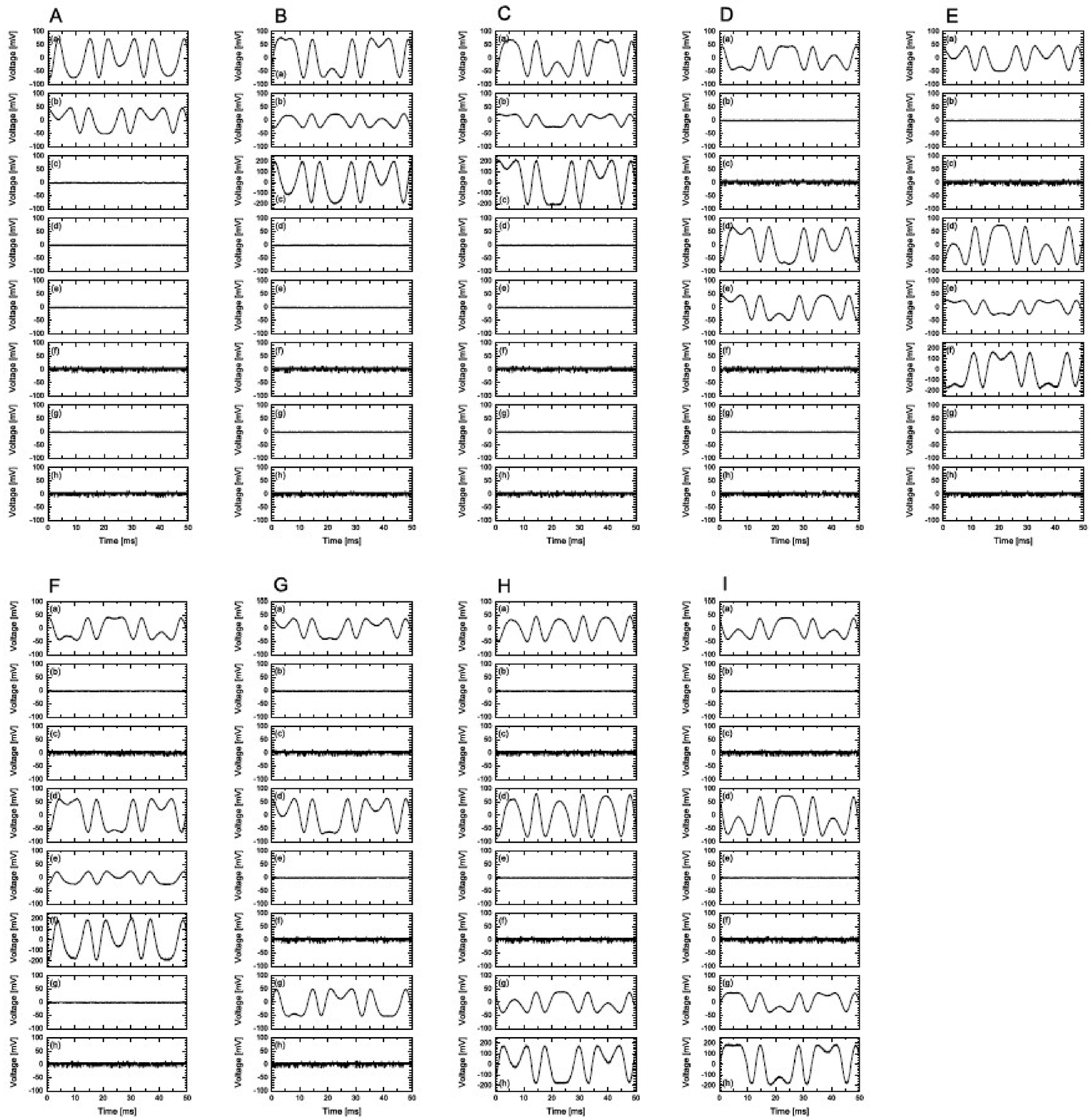}
% \begin{tabular}{ccccc}
% \includegraphics[clip,scale=0.5]{./suppl-figs/ch1gs2.eps} &
% \includegraphics[clip,scale=0.5]{./suppl-figs/ch2gs2.eps} &
% \includegraphics[clip,scale=0.5]{./suppl-figs/ch3gs2.eps} &
% \includegraphics[clip,scale=0.5]{./suppl-figs/ch4gs2.eps} &
% \includegraphics[clip,scale=0.5]{./suppl-figs/ch5gs2.eps} \\
% \includegraphics[clip,scale=0.5]{./suppl-figs/ch6gs2.eps} &
% \includegraphics[clip,scale=0.5]{./suppl-figs/ch7gs2.eps} &
% \includegraphics[clip,scale=0.5]{./suppl-figs/ch8gs2.eps} &
% \includegraphics[clip,scale=0.5]{./suppl-figs/ch9gs2.eps} &
% \end{tabular}
\caption
{Results of error syndrome measurements. A-I correspond to the cases of
an error in quantum channels 1-9, respectively. (a)-(h) correspond to
outputs from homodyne detectors 1-8, respectively.
}\label{fig-syndrome-results}
\end{figure*}

Fig.~\ref{fig-syndrome-results}
shows error syndrome measurement results.
Here, the input state is a vacuum state. This case is also described in the main body of the paper.
A random displacement error in phase space is
transferred onto quantum channels one through nine for A-I, respectively.
With Table 1 in the main body of the paper and Fig.~\ref{fig-syndrome-results} here,
we can decide which quantum channel is subject to an error
and derive a corresponding feedforward operation to correct the error.

\section{Results of error correction}

\begin{figure*}[t]
\centering
\includegraphics[clip,scale=0.5]{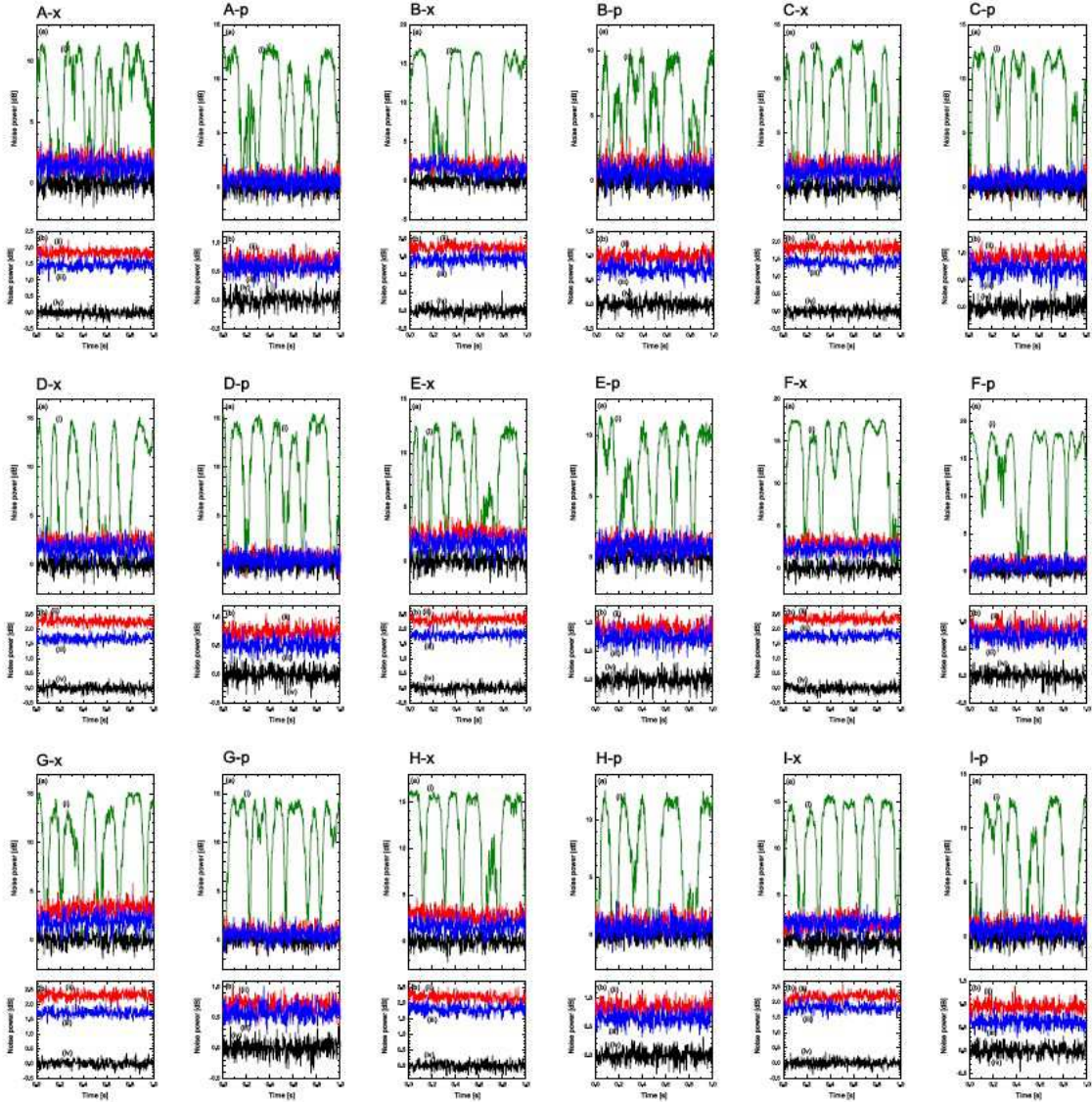}
% \begin{tabular}{cccccc}
% \includegraphics[clip,scale=0.4]{./suppl-figs/ch1xgs2.eps} &
% \includegraphics[clip,scale=0.4]{./suppl-figs/ch1pgs2.eps} &
% \includegraphics[clip,scale=0.4]{./suppl-figs/ch2xgs2.eps} &
% \includegraphics[clip,scale=0.4]{./suppl-figs/ch2pgs2.eps} &
% \includegraphics[clip,scale=0.4]{./suppl-figs/ch3xgs2.eps} &
% \includegraphics[clip,scale=0.4]{./suppl-figs/ch3pgs2.eps} \\
% \includegraphics[clip,scale=0.4]{./suppl-figs/ch4xgs2.eps} &
% \includegraphics[clip,scale=0.4]{./suppl-figs/ch4pgs2.eps} &
% \includegraphics[clip,scale=0.4]{./suppl-figs/ch5xgs2.eps} &
% \includegraphics[clip,scale=0.4]{./suppl-figs/ch5pgs2.eps} &
% \includegraphics[clip,scale=0.4]{./suppl-figs/ch6xgs2.eps} &
% \includegraphics[clip,scale=0.4]{./suppl-figs/ch6pgs2.eps} \\
% \includegraphics[clip,scale=0.4]{./suppl-figs/ch7xgs2.eps} &
% \includegraphics[clip,scale=0.4]{./suppl-figs/ch7pgs2.eps} &
% \includegraphics[clip,scale=0.4]{./suppl-figs/ch8xgs2.eps} &
% \includegraphics[clip,scale=0.4]{./suppl-figs/ch8pgs2.eps} &
% \includegraphics[clip,scale=0.4]{./suppl-figs/ch9xgs2.eps} &
% \includegraphics[clip,scale=0.4]{./suppl-figs/ch9pgs2.eps}
% \end{tabular}
\caption
{Results of error correction. A-I correspond to the cases of an error in quantum channels 1-9, respectively. The LO phase of the homodyne detector is locked at $x$ or $p$, which is indicated after the capital letters.
Trace numbers are the same as in Figs.~(5) and (6) in the main body of the paper.
}\label{fig-correction-results}
\end{figure*}

Fig.~\ref{fig-correction-results} shows the results of error correction.
The results are summarized in Table 2 in the main body of the paper.

\section{The role of multipartite entanglement}

The encoded nine-mode state, as created in the current experiment and described by
eqs.~(\ref{eq-encode-all}), approaches the following state
in the limit of infinite squeezing,
\begin{eqnarray}\label{complete9codeexperiment}
|\psi_{\rm
encode}\rangle&=&\frac{1}{\pi^{3/2}}\int\,dp\,\,dp_1\,dp_2\,dp_3\,\bar
\psi(p)\,e^{-2ip(p_1+p_2+p_3)}
\nonumber\\
&&\times|p_1,p_1,p_1,p_2,p_2,p_2,p_3,p_3,p_3\rangle.
\end{eqnarray}
Clearly, even for infinite squeezing and perfect encoding,
the inseparability properties of the total nine-party state
depend on the signal input wave function $\bar\psi(p)$. In particular,
for $\bar\psi(p)\equiv \delta(p)$, we obtain $|\psi_{\rm
encode}\rangle = \int\,dp\,|p,p,p\rangle \otimes
\int\,dp\,|p,p,p\rangle \otimes \int\,dp\,|p,p,p\rangle$, which is
clearly not fully nine-party entangled, but rather a product
state of three fully tripartite entangled GHZ-type three-mode
states. So in order to obtain full nine-party entanglement, the input state
should not correspond to an infinitely $x$-squeezed state
(corresponding, after Fourier transform, to $\bar\psi(p)\equiv
\delta(p)$). Similarly, an infinitely $p$-squeezed
input state leads to vanishing GHZ-type correlations {\it
within} each of the three triplets, but it has excellent GHZ-type
correlations {\it between} the three triplets. For an input state
between these two extremes, for instance, a vacuum input state as
used in the experiment, we obtain quadrature correlations of
both types, potentially leading to full nine-party entanglement.

In order to witness full nine-party entanglement, in addition to the
correlations of eqs.~(\ref{stabilizer}),
$p$-correlations {\it between} the triplets and
$x$-correlations {\it within} each triplett are required.
Equations~(\ref{stabilizer}) only describe $x$-correlations between the triplets
and $p$-correlations within each triplet. The missing correlations
are of the type of $\hat p_1 + \hat p_4 + \hat p_7 \to 0$ and
$\hat x_1 + \hat x_2 + \hat x_3 \to 0$. These linear combinations
correspond to the ``logical'' quadratures in the code space,
\begin{eqnarray}\label{xcorr_withintriplets}
\hat X &\equiv& \hat x_1 + \hat x_2 + \hat x_3 = \hat x_{\rm in} + \sqrt{2} \hat
x_{\rm an1}^{(0)} e^{-r_1}\,,\nonumber\\
\hat P &\equiv& \hat p_1 + \hat p_4 + \hat p_7 = \hat p_{\rm in} +
\sqrt{\frac{2}{3}} \hat p_{\rm an2}^{(0)} e^{-r_2}\nonumber\\
&& + \sqrt{\frac{2}{3}} \hat p_{\rm an5}^{(0)} e^{-r_5}+
\sqrt{\frac{2}{3}} \hat p_{\rm an7}^{(0)} e^{-r_7}\,,
\end{eqnarray}
obviously depending on the signal input state. Only for infinite ancilla squeezing,
the encoding is perfect, $\hat X = \hat x_{\rm in}$ and $\hat P = \hat p_{\rm in}$.
The excess noise in each quadrature is $2 \times
e^{-2r}/4$ for equal squeezing of the ancilla modes. In this
case, an infinitely $x$-squeezed input state would lead to
excellent intra-triplet $x$-correlations; an infinitely
$p$-squeezed input state, favorable for good inter-triplet
$p$-correlations, leads to vanishing intra-triplet
$x$-correlations. With a vacuum input state, as used in the experiment,
we have both types of
quantum correlations for nonzero squeezing of the ancilla modes.
Similar quantum correlations also exist for the combinations
\begin{eqnarray}\label{xpcorr_2}
\hat x_4 + \hat x_5 + \hat x_6 &=& \hat x_{\rm in} -
\frac{1}{\sqrt{2}} \hat x_{\rm an1}^{(0)} e^{-r_1} +
\sqrt{\frac{3}{2}} \hat x_{\rm an4}^{(0)} e^{-r_4}\,,\nonumber\\
\hat x_7 + \hat x_8 + \hat x_9 &=& \hat x_{\rm in} -
\frac{1}{\sqrt{2}} \hat x_{\rm an1}^{(0)} e^{-r_1} -
\sqrt{\frac{3}{2}} \hat x_{\rm an4}^{(0)} e^{-r_4}\,,\nonumber\\
\hat p_2 + \hat p_5 + \hat p_8 &=& \hat p_{\rm in} -
\frac{1}{\sqrt{6}} \hat p_{\rm an2}^{(0)} e^{-r_2} +
\sqrt{\frac{1}{2}} \hat p_{\rm an3}^{(0)} e^{-r_3} \nonumber\\
&&-
\frac{1}{\sqrt{6}} \hat p_{\rm an5}^{(0)} e^{-r_5} +
\sqrt{\frac{1}{2}} \hat p_{\rm an6}^{(0)} e^{-r_6} \nonumber\\
&&-
\frac{1}{\sqrt{6}} \hat p_{\rm an7}^{(0)} e^{-r_7} +
\sqrt{\frac{1}{2}} \hat p_{\rm an8}^{(0)} e^{-r_8}\,,\nonumber\\
\hat p_3 + \hat p_6 + \hat p_9 &=& \hat p_{\rm in} -
\frac{1}{\sqrt{6}} \hat p_{\rm an2}^{(0)} e^{-r_2} -
\sqrt{\frac{1}{2}} \hat p_{\rm an3}^{(0)} e^{-r_3} \nonumber\\
&&-
\frac{1}{\sqrt{6}} \hat p_{\rm an5}^{(0)} e^{-r_5} -
\sqrt{\frac{1}{2}} \hat p_{\rm an6}^{(0)} e^{-r_6} \nonumber\\
&&-
\frac{1}{\sqrt{6}} \hat p_{\rm an7}^{(0)} e^{-r_7} -
\sqrt{\frac{1}{2}} \hat p_{\rm an8}^{(0)} e^{-r_8}\,.\nonumber\\
\end{eqnarray}

The total set of quadrature quantum correlations can be sufficient for
a fully inseparable nine-party entangled state. The corresponding
nine-party entanglement witnesses lead to the known criteria
for multi-party inseparability of continuous-variable states \cite{vanLoock03}.
In order to verify three-party
inseparability within each triplett $\hat\rho_{123}$,
$\hat\rho_{456}$, and $\hat\rho_{789}$, we have, for $\hat\rho_{123}$,
\begin{eqnarray}\label{criteria_withintriplets1}
\langle[\Delta(\hat{p}_1-\hat{p}_2)]^2\rangle +
\langle[\Delta(\hat{x}_1+\hat{x}_2+g_{1a}\,\hat{x}_3)]^2\rangle <
1,
\nonumber\\
\langle[\Delta(\hat{p}_2-\hat{p}_3)]^2\rangle +
\langle[\Delta(g_{1b}\,\hat{x}_1+\hat{x}_2+\hat{x}_3)]^2\rangle <
1,
\nonumber\\
\end{eqnarray}
for $\hat\rho_{456}$,
\begin{eqnarray}\label{criteria_withintriplets2}
\langle[\Delta(\hat{p}_4-\hat{p}_5)]^2\rangle
+ \langle[\Delta(\hat{x}_4+\hat{x}_5+g_{2a}\,\hat{x}_6)]^2\rangle
< 1,
\nonumber\\
\quad \langle[\Delta(\hat{p}_5-\hat{p}_6)]^2\rangle
+ \langle[\Delta(g_{2b}\,\hat{x}_4+\hat{x}_5+\hat{x}_6)]^2\rangle
< 1,
\nonumber\\
\end{eqnarray}
for $\hat\rho_{789}$,
\begin{eqnarray}\label{criteria_withintriplets3}
\langle[\Delta(\hat{p}_7-\hat{p}_8)]^2\rangle
+
\langle[\Delta(\hat{x}_7+\hat{x}_8+g_{3a}\,\hat{x}_9)]^2\rangle
< 1,
\nonumber\\
\langle[\Delta(\hat{p}_8-\hat{p}_9)]^2\rangle
+
\langle[\Delta(g_{3b}\,\hat{x}_7+\hat{x}_8+\hat{x}_9)]^2\rangle
< 1.
\nonumber\\
\end{eqnarray}
The ``gains'' $g_{1a}$, etc., can be used to optimize these
conditions. In order to rule out a state of the form $\sum_i
\eta_i\hat\rho_{123}^{(i)} \otimes \hat\rho_{456}^{(i)} \otimes
\hat\rho_{789}^{(i)} \equiv \sum_i\eta_i\hat\rho_a^{(i)} \otimes
\hat\rho_b^{(i)} \otimes \hat\rho_c^{(i)}$, we need further
criteria, for example,
\begin{eqnarray}\label{criteria_betweentriplets1}
\langle[\Delta(\hat{p}_1+\hat{p}_4+g_{11}\,\hat{p}_7)]^2\rangle
\quad\quad\quad\quad\quad\quad\quad\quad\quad\quad\quad\quad
\quad\quad\;\\
+ \langle[\Delta(\hat{x}_1+g_{12}\,\hat{x}_2+g_{13}\,\hat{x}_3-
\hat{x}_4-g_{14}\,\hat{x}_5-g_{15}\,\hat{x}_6)]^2\rangle <
1\nonumber\\
\nonumber\\\label{criteria_betweentriplets2}
\langle[\Delta(g_{21}\,\hat{p}_1+\hat{p}_4+\hat{p}_7)]^2\rangle
\quad\quad\quad\quad\quad\quad\quad\quad\quad\quad\quad\quad
\quad\quad\;\\
+ \langle[\Delta(\hat{x}_4+g_{22}\,\hat{x}_5+g_{23}\,\hat{x}_6-
\hat{x}_7-g_{24}\,\hat{x}_8-g_{25}\,\hat{x}_9)]^2\rangle < 1
\nonumber
\end{eqnarray}
which describe the inter-triplet correlations.
Equation~(\ref{criteria_betweentriplets1}) rules
out the forms $\sum_i\eta_i\hat\rho_a^{(i)} \otimes
\hat\rho_{bc}^{(i)}$ and $\sum_i\eta_i\hat\rho_b^{(i)} \otimes
\hat\rho_{ac}^{(i)}$; eq.~(\ref{criteria_betweentriplets2})
rules out the forms
$\sum_i\eta_i\hat\rho_c^{(i)} \otimes \hat\rho_{ab}^{(i)}$ and
$\sum_i\eta_i\hat\rho_b^{(i)} \otimes \hat\rho_{ac}^{(i)}$. Thus,
any form of separability between the triplets $a$, $b$, and $c$
can be ruled out. The
inter-triplet conditions can be understood as GHZ-type
correlations of modes 1, 4, and 7 after LOCC operations; namely,
$x$-measurements of modes 2, 3, 5, 6, 8, 9 and the corresponding
displacements of modes 1, 4, and 7.

In the experiment, it was verified that
in any of the nine cases of an error in any one
of the nine channels,
the classical cutoff (zero-squeezing limit) was exceeded.
This confirms
that all 8 ancilla modes are in a squeezed state
(see Table 1 and Fig.2), as the
quadrature noise of every ancilla mode contributes to the excess noise
of the corrected signal for some of the detector results used for
feedforward. This squeezing translates into nonclassical correlations
for all combinations in eqs.~(\ref{eq-th2-4}),
(\ref{xcorr_withintriplets}), and (\ref{xpcorr_2}) (with a vacuum
input state). The set of quadrature combinations corresponds to the
``unit-gain'' version of the entanglement witnesses in
eqs.~(\ref{criteria_withintriplets1}), (\ref{criteria_withintriplets2}),
(\ref{criteria_withintriplets3}),
(\ref{criteria_betweentriplets1}), and (\ref{criteria_betweentriplets2}).
In order to satisfy the witness inequalities, in particular, for small
squeezing values (as those of roughly 1 dB in the experiment),
non-unit gain must be chosen.
Although these non-unit gain combinations have not been measured directly
in the quantum error correction experiment, the nonclassicality in {\it all} the
unit-gain combinations may be interpreted as an indirect confirmation of the
presence of nine-party entanglement.

\section{Applicability of continuous-variable codes}

The continuous-variable nine-mode code corrects an arbitrary
error occurring in {\it any one} of the nine channels. Similar
to the qubit case, for realistic scenarios, we should consider
imperfect transmissions in {\it every} channel under the reasonable
assumption of errors acting independently in all the channels.
The performance of the code can then be evaluated by comparing the
transfer fidelities for the encoded scheme with a direct transmission
of the signal state through a single noisy channel \cite{NielsenChuang}.

Using the example of a 3-wavepacket code, we shall demonstrate
that for certain stochastic error models, the continuous-variable
code leads to a dramatic improvement of fidelity even when the
errors occur in every channel \cite{PvL08}. 
In this case, the errors should
correspond to $x$-displacements or any errors decomposable into
$x$-displacements (including non-Gaussian ``$x$-errors'').
A code for correcting arbitrary errors including non-commuting
$x$ and $p$-errors is obtainable, for instance, by concatenating the
3-mode code into a 9-mode code, as implemented in the current experiment.
The appropriate error models are reminiscent of the most typical qubit
channels such as bit-flip and phase-flip channels.
In the continuous-variable regime,
these types of stochastic errors would map a Gaussian signal state
into a non-Gaussian state represented by a discrete, incoherent mixture of
the input state with a Gaussian (or even a non-Gaussian) state,
\begin{eqnarray}\label{errormodel}
W_{\rm out}(x,p) = (1-\gamma)  W_{\rm in} + \gamma W_{\rm error}\,.
\end{eqnarray}
Here, the input state described
by the Wigner function $W_{\rm in}$ is transformed into a new state
$W_{\rm error}$
with probability $\gamma$; it remains unchanged with probability
$1-\gamma$. A special case of the above channel model is an
erasure channel \cite{Niset}. The generalized erasure model here
may find applications in free-space communication with fluctuating
losses and beam point jitter effects \cite{Heersink,Dong,Schnabel}.

As an example, we will consider a coherent-state input,
$|\bar\alpha_1\rangle = |\bar x_1 + i \bar p_1\rangle$,
described by the Wigner function,
\begin{eqnarray}\label{input}
W_{\rm in}(x_1,p_1) = \frac{2}{\pi} \exp[-2 (x_1 - \bar x_1)^2
-2 (p_1 - \bar p_1)^2]\,.
\end{eqnarray}
Moreover, we assume that the effect of the error
is just an $x$-displacement by $\bar x_2$ such that
\begin{eqnarray}\label{output}
W_{\rm error}(x_1,p_1) = W_{\rm in}(x_1-\bar x_2,p_1)\,.
\end{eqnarray}
The sign of the displacement error is fixed and known,
e.g., without loss of generality, $\bar x_2 > 0$.
Note that more general errors, including non-Gaussian $x$-errors,
could be considered as well.

Now in order to encode the input state, we use
two ancilla modes, each in a single-mode $x$-squeezed vacuum state, represented by
\begin{eqnarray}\label{ancilla}
W_{\rm anc}(x_k,p_k) = \frac{2}{\pi} \exp[-2 e^{+2 r} x_k^2
-2 e^{-2 r} p_k^2]\,,
\end{eqnarray}
with squeezing parameter $r$ and $k=2,3$.
The total three-mode state before encoding is
\begin{eqnarray}\label{totalbefore}
W(\alpha_1,\alpha_2,\alpha_3) = W_{\rm in}(x_1,p_1)
W_{\rm anc}(x_2,p_2)W_{\rm anc}(x_3,p_3),\nonumber\\
\end{eqnarray}
with $\alpha_j = x_j + i p_j$, $j=1,2,3$. The encoding may be achieved
by applying a ``tritter'', i.e., a sequence of two beam splitters with
transmittances $1:2$ and $1:1$. The total, encoded state will be an
entangled three-mode Gaussian state with Wigner function,
\begin{eqnarray}\label{totalafter}
&&W_{\rm enc}(\alpha_1,\alpha_2,\alpha_3) =
\left(\frac{2}{\pi}\right)^3 \\
&&\times
\exp\Big\{-2 \Big[\frac{1}{\sqrt{3}} \Big(x_1 + x_2 + x_3\Big) - \bar x_1\Big]^2
\nonumber\\
&&\quad\quad\quad\,\,\,
-\frac{2}{3} e^{-2 r} \Big[(p_1-p_2)^2 + (p_2-p_3)^2 + (p_1-p_3)^2\Big]
\nonumber\\
&&\quad\quad\quad\,\,\,
-2 \Big[\frac{1}{\sqrt{3}} \Big(p_1 + p_2 + p_3\Big) - \bar p_1\Big]^2
\nonumber\\
&&\quad\quad\quad\,\,\,
-\frac{2}{3} e^{+2 r} \Big[(x_1-x_2)^2 + (x_2-x_3)^2 + (x_1-x_3)^2\Big]\Big\}.
\nonumber
\end{eqnarray}
Now we send the three modes through individual channels where each channel
acts independently upon {\it every} mode as described by Eq.~(\ref{errormodel})
with $W_{\rm error}$ corresponding to an $x$-displacement by $\bar x_2$.
As a result, the three noisy channels will turn the encoded state into
the following three-mode state,
\begin{eqnarray}\label{totalafterchannel}
&&W_{\rm enc}'(\alpha_1,\alpha_2,\alpha_3) \\
&&=(1-\gamma)^3
W_{\rm enc}(\alpha_1,\alpha_2,\alpha_3)
\nonumber\\
&&\quad + \gamma (1-\gamma)^2
W_{\rm enc}(x_1-\bar x_2 + i p_1,\alpha_2,\alpha_3)
\nonumber\\
&&\quad + \gamma (1-\gamma)^2
W_{\rm enc}(\alpha_1,x_2-\bar x_2 + i p_2,\alpha_3)
\nonumber\\
&&\quad + \gamma (1-\gamma)^2
W_{\rm enc}(\alpha_1,\alpha_2,x_3-\bar x_2 + i p_3)
\nonumber\\
&&\quad + \gamma^2 (1-\gamma)
W_{\rm enc}(x_1-\bar x_2 + i p_1,x_2-\bar x_2 + i p_2,\alpha_3)
\nonumber\\
&&\quad + \gamma^2 (1-\gamma)
W_{\rm enc}(x_1-\bar x_2 + i p_1,\alpha_2,x_3-\bar x_2 + i p_3)
\nonumber\\
&&\quad + \gamma^2 (1-\gamma)
W_{\rm enc}(\alpha_1,x_2-\bar x_2 + i p_2,x_3-\bar x_2 + i p_3)
\nonumber\\
&&\quad + \gamma^3
W_{\rm enc}(x_1-\bar x_2 + i p_1,x_2-\bar x_2 + i p_2,x_3-\bar x_2 + i p_3).
\nonumber
\end{eqnarray}
Note that we assumed the same $x$-displacements in every channel.

The decoding procedure now simply means inverting the tritter, which results in
\begin{eqnarray}\label{totalafterdecoding}
&&W_{\rm dec}(\alpha_1,\alpha_2,\alpha_3) \\
&&=(1-\gamma)^3
W_{\rm in}(x_1,p_1)W_{\rm anc}(x_2,p_2)W_{\rm anc}(x_3,p_3)
\nonumber\\
&&\quad + \gamma (1-\gamma)^2
W_{\rm in}\left(x_1-\frac{1}{\sqrt{3}}\bar x_2,p_1\right)\nonumber\\
&&\quad\quad\times
W_{\rm anc}\left(x_2-\sqrt{\frac{2}{3}}\bar x_2,p_2\right)
W_{\rm anc}(x_3,p_3)
\nonumber\\
&&\quad + \gamma (1-\gamma)^2
W_{\rm in}\left(x_1-\frac{1}{\sqrt{3}}\bar x_2,p_1\right)\nonumber\\
&&\quad\quad\times
W_{\rm anc}\left(x_2+\frac{1}{\sqrt{6}}\bar x_2,p_2\right)
W_{\rm anc}(x_3-\frac{1}{\sqrt{2}}\bar x_2,p_3)
\nonumber\\
&&\quad + \gamma (1-\gamma)^2
W_{\rm in}\left(x_1-\frac{1}{\sqrt{3}}\bar x_2,p_1\right)\nonumber\\
&&\quad\quad\times
W_{\rm anc}\left(x_2+\frac{1}{\sqrt{6}}\bar x_2,p_2\right)
W_{\rm anc}(x_3+\frac{1}{\sqrt{2}}\bar x_2,p_3)
\nonumber\\
&&\quad + \gamma^2 (1-\gamma)
W_{\rm in}\left(x_1-\frac{2}{\sqrt{3}}\bar x_2,p_1\right)\nonumber\\
&&\quad\quad\times
W_{\rm anc}\left(x_2-\frac{1}{\sqrt{6}}\bar x_2,p_2\right)
W_{\rm anc}(x_3-\frac{1}{\sqrt{2}}\bar x_2,p_3)
\nonumber\\
&&\quad + \gamma^2 (1-\gamma)
W_{\rm in}\left(x_1-\frac{2}{\sqrt{3}}\bar x_2,p_1\right)\nonumber\\
&&\quad\quad\times
W_{\rm anc}\left(x_2-\frac{1}{\sqrt{6}}\bar x_2,p_2\right)
W_{\rm anc}(x_3+\frac{1}{\sqrt{2}}\bar x_2,p_3)
\nonumber\\
&&\quad + \gamma^2 (1-\gamma)
W_{\rm in}\left(x_1-\frac{2}{\sqrt{3}}\bar x_2,p_1\right)\nonumber\\
&&\quad\quad\times
W_{\rm anc}\left(x_2+\sqrt{\frac{2}{3}}\bar x_2,p_2\right)
W_{\rm anc}(x_3,p_3)
\nonumber\\
&&\quad + \gamma^3
W_{\rm in}\left(x_1-\sqrt{3}\bar x_2,p_1\right)\nonumber\\
&&\quad\quad\times
W_{\rm anc}\left(x_2,p_2\right)
W_{\rm anc}(x_3,p_3).
\nonumber
\end{eqnarray}
By looking at this state, we can easily see that
$x$-homodyne detections of the ancilla modes 2 and 3
(the syndrome measurements) will
almost unambiguously identify in which channel
a displacement error occurred and
how many modes were subject to a displacement error.
The only ambiguity comes from the case of an error
occurring in every channel at the same time (with probability $\gamma^3$),
which is indistinguishable from the case where no error at all happens.
In both cases, the two ancilla modes
are transformed via decoding back into the two initial
single-mode squeezed vacuum states. All the other cases, however,
can be identified, provided the initial squeezing $r$ is sufficiently
large such that the displacements $\propto\bar x_2$, originating from
the errors, can be resolved in the ancilla states.

The recovery operation, i.e., the final phase-space displacement
of mode 1 depends on the syndrome measurement results for modes 2 and 3
which are consistent with either undisplaced squeezed vacuum states
(`$0$') or squeezed vacua displaced in either `$+$' or `$-$' $x$-direction.
The syndrome results for modes 2 and 3 corresponding to the
eight possibilities for the errors occurring in the three channels
are ($0$,$0$) for no error at all, ($+$,$0$) for an error in channel 1,
($-$,$+$) for an error in channel 2, ($-$,$-$) for an error in channel 3,
($+$,$+$) for errors in channels 1 and 2, ($+$,$-$) for errors in channels 1 and 3,
($-$,$0$) for errors in channels 2 and 3, and, again, ($0$,$0$) for errors
occurring in all three channels.

In the limit of infinite squeezing of the ancilla modes,
the ensemble output state of mode 1
(upon averaging over all syndrome measurement results
$x_2$ and $x_3$ including suitable feedforward operations)
can be described as
\begin{eqnarray}\label{ensembleoutput}
(1-\gamma^3)W_{\rm in}(x_1,p_1) +
\gamma^3 W_{\rm in}\left(x_1-\sqrt{3}\bar x_2,p_1\right)\,.
\end{eqnarray}
This output state emerges, because
in almost all cases, the feedforward operations
turn mode 1 back into the initial state (in the case of finite squeezing,
only up to some Gaussian-distributed excess
noise depending on the degree of squeezing used for the encoding).
The only case for which no correction occurs is when errors appear
in every channel at the same time, at a probability of $\gamma^3$.
In this case, the initial state remains uncorrected, with an $x$-displacement
error of $\sqrt{3}\bar x_2$.

We see that a fidelity of $1-\gamma^3$ can be achieved,
assuming $\bar x_2\gg 1$ (for smaller $\bar x_2$, the fidelity
would even exceed $1-\gamma^3$, but those smaller $\bar x_2$ may be too
hard to detect at the syndrome extraction, depending on the
degree of squeezing, see below).
Note that this result implies that the encoded scheme performs
better than the unencoded scheme (direct transmission with
$F_{\rm direct}=1-\gamma$) for {\it any} $0<\gamma<1$.
In other words, by employing the quantum error correction protocol,
the error probability can be reduced from
$\gamma$ to $\gamma^3$. The continuous-variable scheme
in this model is more efficient than the analogous qubit repetition
code and it does not require error probabilities $\gamma < 1/2$
as for the case of qubit bit-flip errors \cite{NielsenChuang}.

We may now consider two different regimes for the error displacements
$\bar x_2$. First, the regime $e^{-2r}/4 < \bar x_2 < 1/4$, corresponding
to small displacements below the shot noise limit;
these can only be resolved provided
the squeezing is large enough. In the limit of infinite squeezing
$r\to\infty$, arbitrarily small shifts can be detected and
perfectly corrected (with zero excess noise in the output states
corresponding to unit fidelity).

Secondly, the regime $\bar x_2\gg 1$.
For these large shifts, even zero squeezing in the ancilla modes
(i.e., vacuum ancilla states) is sufficient for error identification.
Even with $r=0$, the syndrome measurements still provide enough
information on the location of the error and, to some extent, on the size
of the error. We may refer to this kind of scheme as
classical error correction (CEC), corresponding to the ``classical cutoff''
used as a classical boundary in the main body of the paper.
This classical cutoff depends on the particular encoding
and decoding circuit used; in the experiment, it is the same circuit
as that employed for quantum error correction (neither of these are
necessarily optimal).

CEC for large
shifts $\bar x_2\gg 1$ (the regime of the experiment) works fairly well.
In fact, the fidelity values without CEC drop
to near-zero fidelities, as measured in the experiment, $F<0.007\pm0.001$.
Experimentally, this CEC is a highly nontrivial task
and it is needed to achieve reasonable transfer fidelities.
Nonetheless, the CEC scheme results in excess noise for the output state
coming from the feedforward operations based on the fluctuating
syndrome measurement results. By employing squeezed-state ancilla
modes, this excess noise can be reduced (down to zero for infinite
squeezing). In this case, the scheme operates in the quantum regime.
Significantly, in the experiment, non-commuting errors have been corrected,
which means that CEC will always result in some excess noise.
Although the margin of the demonstrated quantum error correction
(on top of the CEC) is rather small, the experimental data provide
clear evidence that CEC has been outperformed by the quantum scheme.


\begin{thebibliography}{99}
\bibitem{nogo} note that it was proven recently that
quantum error correction of Gaussian errors for
Gaussian signal states using
Gaussian resources and Gaussian operations is impossible
\cite{Nisetqph};
the error models referred to here 
\cite{PvL08}
and discussed in the appendix correspond to non-Gaussian errors.
\bibitem{footnote}
in the experiment, we generated a code state
with position $x$ and momentum $p$ interchanged.
% $|\psi_{\rm encode}\rangle= \int\,dX\,
% \bar\psi(X)\,|X,X,X\rangle$
% with $|X\rangle = \frac{1}{\sqrt{\pi}}\int\,dp\,
% \,e^{-2iX p}\,|p,p,p\rangle$.
% Again, an arbitrary single-mode state $|\psi\rangle
% =\int\,dp\,\bar\psi(p)|p\rangle$ is encoded into nine optical modes.
This encoding (and the corresponding QEC protocol) involves only a change
of basis with no drop in performance. Quantum optically, this change
corresponds to a 90-degree rotation of the quadrature amplitudes,
requiring local oscillator phases to be shifted by 90 degrees for homodyne
detection.
\end{thebibliography}
\end{document}